\documentclass[a4paper,12pt]{article}

\title{Bond Graphs Unify Stoichiometric Analysis\\ and Thermodynamics}
\usepackage[noblocks]{authblk}
\author[1,2]{Peter J. Gawthrop\footnote{Corresponding
    author. \textbf{peter.gawthrop@unimelb.edu.au}}}
\affil[1]{
  Systems Biology Laboratory,
  Department of Biomedical Engineering,
  Melbourne School of Engineering,
  University of Melbourne,
  Victoria 3010, Australia.
   }
   \affil[2]{Systems Biology Laboratory,
     School of Mathematics and Statistics,
     University of
      Melbourne University of Melbourne, Victoria 3010, Australia}

\usepackage{abstract}

\usepackage{a4wide,times,siunitx}
\usepackage{framed}

\usepackage[authoryear,round]{natbib}
\usepackage[hidelinks]{hyperref}
\usepackage{url,doi}
\usepackage[section]{placeins}

\newcommand{\Ref}[1]{\ref{#1}, p.\pageref{#1}}
\newcommand{\cref}[1]{Chapter~\Ref{#1}}

\newcommand{\BG}[1]{\text{\sffamily\textbf{#1}}}

\newcommand{\Ce}{\BG{Ce }}

\newcommand{\one}{\BG{1 }}
\newcommand{\zero}{\BG{0 }}

\renewcommand{\Re}{\BG{Re }}

\newcommand{\BGL}[2]{$\BG{#1}$:$\mathbf{#2}$} %Generic

\newcommand{\BCe}[1]{\BGL{Ce}{#1}}

\newcommand{\bgbond}{ {$\rightharpoondown$} }

\usepackage{amsmath,amssymb,amscd,amstext,mathtools,extarrows,centernot}
\numberwithin{equation}{section}

\newcommand{\lb}{\left (}
\newcommand{\rb}{\right )}
\newcommand{\diag}{\text{diag}}
\newcommand{\where}{\text{where }}

\newcommand{\NN}{N}
\newcommand{\Ncd}{\NN^{cd}}
\newcommand{\Nf}{{\NN^f}}
\newcommand{\Nr}{{\NN^r}}
\newcommand{\Np}{{\NN_p}}

\newcommand{\KK}{K}
\newcommand{\Kcd}{\KK^{cd}}

\newcommand{\xx}{x}

\newcommand{\Phif}{{\Phi^f}}
\newcommand{\Phir}{{\Phi^r}}

\newcommand{\phis}{{\phi^\std}}

\newcommand{\vv}{v}

\usepackage{listings}
\usepackage{chemformula}

\newcommand{\std}{\oslash}

\newcommand{\Ch}[1]{(\ch{#1})}

\usepackage{graphicx,subfigure,fancybox,color}
\graphicspath{{./Figs/}}

\newcommand{\Fig}[2]{
 \includegraphics[width=#2\linewidth]{#1.pdf}
}

\newcommand{\SubFig}[3]{
 \subfigure[#2]{
   \includegraphics[width=#3\linewidth]{#1.pdf}
   \label{subfig:#1}
 }
}

\newcommand{\potential}{free energy }
\newcommand{\potentials}{free energies }
\newcommand{\Pathway}[4]{
  This gives rise to the pathway:
\begin{itemize}
  #3
\end{itemize}
The corresponding pathway reaction is:
\begin{xalignat*}{2}
   #4
\end{xalignat*}
  }

\begin{document}
\maketitle
\begin{abstract}
Whole-cell modelling is constrained by the laws of nature in
general and the laws of thermodynamics in particular.  This paper
shows how one prolific source of information, stoichiometric models
of biomolecular systems, can be integrated with thermodynamic
principles using the bond graph approach to network thermodynamics.
\end{abstract}
\newpage
\tableofcontents
\newpage
\section{Introduction}
Whole-cell modelling has the potential to ``predict phenotype from
genotype''~\citep{KarSanMac12,Cov15} and has the potential to
``transform bioscience and medicine''~\citep{SziYosSek18}. However,
there are currently significant issues in achieving reproducibility
\citep{MedGolKar16} and integrating disparate sources of information
\citep{GolSziChe18}.
However, whatever the source of information, the whole-cell model is
constrained by the laws of nature in general and the laws of
thermodynamics in particular.
Unfortunately, ``The requirement for thermodynamic consistency,
however, has not, in general, been adopted for whole-cell modelling''
\citep{SmiCra04}. This paper shows how one prolific source of
information, stoichiometric models of biomolecular systems, can be
integrated with thermodynamic principles.

Stoichiometric analysis of biomolecular systems has been developed
over the years \citep{HeiSch96,Pal06,Pal11,Pal15} has had notable
successes including modelling and analysis of the E.coli genome-scale
reconstruction \citep{OrtConNa11,ThiSwaFle13,SwaSmaHef16}. The basic idea is to
describe a biomolecular system as a (sparse) integer matrix -- the
$n_X \times n_V$ stoichiometric matrix $\NN$%
\footnote{The stoichiometric matrix has the symbol $\NN$ is some works
  \citep{KliLieWie16} and $S$ in others \citep{Pal06,Pal11,Pal15}}
connecting $n_X$ species and $n_V$ reactions.
As discussed by \citet{Pal15} the stoichiometric approach has a number of
advantages:
\begin{enumerate}
\item The coefficients of $\NN$ are integer; they can therefore be
  determined exactly.
\item Mass balance of species is ensured and, with the inclusion of
  the elemental matrix \citep[\S~9.2.2]{Pal15}, mass balance of
  elements is also ensured.
\item The sparse integer matrix representation is scaleable to include
  large systems; for example, the iJO1366 genome-scale reconstruction
  of the metabolic network of Escherichia coli has 2251 metabolic
  reactions, and 1136 unique metabolites \citep{OrtConNa11}.
\item Standard linear algebraic concepts such as the null spaces of a
  matrix can be invoked to provide precise and meaningful analysis of
  pathways and conserved moieties
  \citep{Pal06,Pal11,Pal15,KliLieWie16}.
\item As discussed by \citet{OrtConNa11}, the flux-balance analysis
  technique \citep{OrtThiPal10} can be applied to predict metabolic
  flux distributions, growth rates, substrate uptake rates, and
  product secretion rates for large models.
\item Because the enzymes catalysing the reactions are related to the
  genome, the stoichiometric approach provides a bridge from genotype
  to phenotype~\citep{Pal15}. 
% \item The null space of the stoichiometric matrix $\NN$, and thus
%   corresponding pathways, are not unique. However, \cite{BorNagLew14}
%   introduce the MinSpan concept and show that ``The MinSpan pathways
%   are the sparsest linear basis of the null space of S that maintains
%   the biological and thermodynamic constraints of the network.''
\item Comprehensive software tools are readily available \citep{EbrLerPal11,HeiArrPfa19}.
\end{enumerate}

A number of works have discussed the fundamental significance of
energy in the life sciences and evolution of living systems
\citep{NivLau08,SouThiLan13,MarSouLan14,Lan14,Lan18,DaiLoc18,NieLeuHei19}.
In particularly, the \emph{efficiency}
\citep{SmiBarLoi05,LopDha14,Niv16,ParRubXu16,LarTorLin16} of living
systems is an evolutionary pressure.
However, energy considerations are not explicitly included in the
stoichiometric approach. This can lead to mass flows that are not
thermodynamically possible; such non-physical flows can be detected
and eliminated by adding additional thermodynamic constraints via
\emph{Energy Balance Analysis} (EBA)
\citep{BeaLiaQia02,QiaBeaLia03,NooBarFla14,Noo18}.

Like living systems, engineering systems are subject to the laws of
physics in general and the laws of thermodynamics in particular.  This
fact gives the opportunity of applying energy-based engineering
approaches to the modelling, analysis and understanding of living
systems.
The bond graph method of \citet{Pay61} is one such well-established
engineering approach~\citep{Cel91,GawSmi96,GawBev07,Bor10,KarMarRos12}
which has been extended to include biomolecular
systems~\citep{OstPerKat71,OstPerKat73,GawCra14}. The stoichiometric
matrix of a biomolecular network can be derived from the corresponding
bond graph \citep{GawCra14,GawCurCra15}; this paper shows that the
converse is true: the bond graph of a biomolecular system can be
deduced from the stoichiometric representation. Thus the large
repository of models of biomolecular systems available in
stoichiometric form can be automatically converted to bond graph
form.

Once converted to bond graph  form, the models are endowed with a
number of additional features:
\begin{enumerate}
\item They are thermodynamically compliant and thus subsume the EBA approach.
\item As an energy based method, bond graphs can model multi-domain
  systems and thus readily incorporate charged species, electrons and
  protons in an integrated
  model \citep{Gaw17a,GawSieKam17,PanGawTra18a,PanGawTra18}.
\item Bond graphs are modular \citep{GawCurCra15,GawCra16} a key
  requirement of any large-scale modelling endeavour.
\item Bond graph models can be simplified in an energetically coherent
  fashion \citep{GawCra14,PanGawCurTraCra17X,GawCudCra19X}.   
\item Bond graphs provide energy-based pathway analysis \citep{GawCra17}.
\end{enumerate}

The \emph{e.coli} Core Model \citep{OrtFlePal10,Pal15} is a
well-documented and readily-available stoichiometric model of a
biomolecular system. This model is used in
\S~\ref{sec:example:-glycolysis-} as an exemplar to illustrate how a
bond graph can be automatically generated and to examine how it can be
used for the energetic analysis of pathways.

\section{Bond Graphs Integrate Stoichiometry and Energy}
\label{sec:bond-graphs-integr}

% \subsection{Bond graphs}\label{sec:bond-graphs}
% \label{sec:bond-graphs-1}
Bond graphs are, as the name implies, a graphical
representation of a system. This has the advantage of clear visual
representation when dealing with small systems, but such
visualisation becomes problematic for large systems. As meaningful
biomolecular systems are large, this issue must be addressed.
There are two approaches to overcoming this issue: modularity and a
non-graphical representation. This paper uses both approaches: a recent
concept of bond graph modularity \citep{Gaw17a} is presented in
\S~\ref{sec:modularity} and the recently developed BondGraphTools
\citep{CudGawPanCra19X} (\url{https://pypi.org/project/BondGraphTools/})
is used throughout as a non-graphical representation.

The key concept is the \emph{energy bond} represented by the \bgbond
symbol. This bond carries energy in the form of an \emph{effort/flow}
pair: in the case of biomolecular systems this pair is chemical
\potential $\phi~\si{\joule\per\mol}$%
\footnote{The symbol $\phi$ is used for chemical \potential in place of
$\mu$.}%
and molar flow
$v~\si{\mol\per\second}$.
Bonds transmit, but do not store or dissipate energy.
Within this context, the bonds connect  four bond graph  components:
\begin{description}
\item[\zero \& \one junctions] Provide a method of connecting a two or
  more bonds. The bonds impinging on a \zero junction share a common
  effort (chemical \potential); the bonds impinging on a \one junction
  share a common flow. Both \zero \& \one junctions transmit, but do
  not store or dissipate energy.
  As discussed previously \citep{GawCra14}, the arrangement of bonds and
  junctions determines the stoichiometry of the corresponding
  biomolecular system and thus the relationship both between reaction and
  species flows and between species \potentials and reaction forward and
  reverse \potentials.
  As will be discussed, the reverse is also true: the
  stoichiometric matrix of a biomolecular system determines the bond
  graph.

\item[\Ce] Represents \emph{species}. Thus species \ch{A} is
  represented by \BCe{A} with the equations:
  \begin{align}
  x_A(t) &= \int_0^t v_A(t^\prime) dt^\prime  + x_A(0)\label{eq:x_A}\\
  \phi_A &= \phi_A^\std + RT \ln \frac{x_A}{x_A^\std} \label{eq:mu_A}
  \end{align}
  Equation (\ref{eq:x_A}) accumulates the flow $v_A$ of species
  \ch{A}.
  Equation (\ref{eq:mu_A}) generates chemical \potential $\phi_A$ in
  terms of the standard \potential $\phi_A^\std$ at standard conditions
  $x_A^\std$ where $R$ and $T$ are the universal gas constant and
  temperature respectively \cite{AtkPau11}. \Ce components store, but
  do not dissipate, energy.
\item[\Re] Represents \emph{reactions}. The flow associated with
  reaction 1 $v_1$ is given by the
\emph{Marcelin -- de Donder} formula~\citep{Rys58}:
  \begin{align}
  v_1 &= \kappa_1 \lb \exp \frac{\Phif_1}{RT} - \exp \frac{\Phir_1}{RT} \rb\label{eq:v_1}
  \end{align}
  where $\Phif_1$ and $\Phir_1$ are the forward and reverse reaction
  \potentials, or affinities. If $\kappa_1$ is constant, this
  represents the mass-action formula; in general, $\kappa_1$ is a
  function of $\Phif_1$, $\Phir_1$ and enzyme
  concentration~\citep{GawCra14}.  \Re components dissipate, but do not
  store, energy. In general
  \begin{align}
    V &= V(\Phi,\phi)\label{eq:V()}\\
    \where
    \Phi &= \Phif - \Phir
  \end{align}
  where $V()$ is dissipative in $\Phi$ for all $\phi$:
  \begin{equation}\label{eq:diss}
    V_i \Phi_i > 0
  \end{equation}
  
\end{description}

% \begin{figure}[htbp]
%   \centering
%   \Fig{Closed_bg}{0.5}
%   \caption{Generic bond graph of closed system}
%   \label{fig:Closed_bg}
% \end{figure}

The key \emph{stoichiometric} equations arising from bond graph
analysis are \citep{GawCra14}:
\begin{align}
  \dot{X} &= NV \label{eq:dX_closed}\\
  \Phi &= -N^T \phi\label{eq:Phi_closed}
\end{align}
where $X$, $\Phi$ and $\phi$ are the species amounts, reaction
\potentials and species \potentials respectively. $N$ is the system
stoichiometric matrix.
The network of bonds and junctions transmits, but does not dissipate
or store, energy. As discussed by \cite{GawCra14}, this fact can be
used to derive Equation~(\ref{eq:Phi_closed}) from
(\ref{eq:dX_closed}). 

Moreover, the stoichiometric matrix $N$ can be decomposed as \citep{GawCra14}:
\begin{equation}
  \label{eq:NfNr}
  \NN = \Nr - \Nf
\end{equation}
where $\Nr$ corresponds to the positive entries of $\NN$ and $\Nf$ to
the negative entries. The forward and reverse reaction \potentials
$\Phif$ and $\Phir$ are given by:
\begin{align}
  \Phif &= \Nf \phi \label{eq:Phif}\\
  \Phir &= \Nr \phi \label{eq:Phir}
\end{align}

% \subsection{From stoichiometry to bond graph}
% \label{sec:from-stoich-bond}
\begin{figure}[htbp]
  \centering
  \SubFig{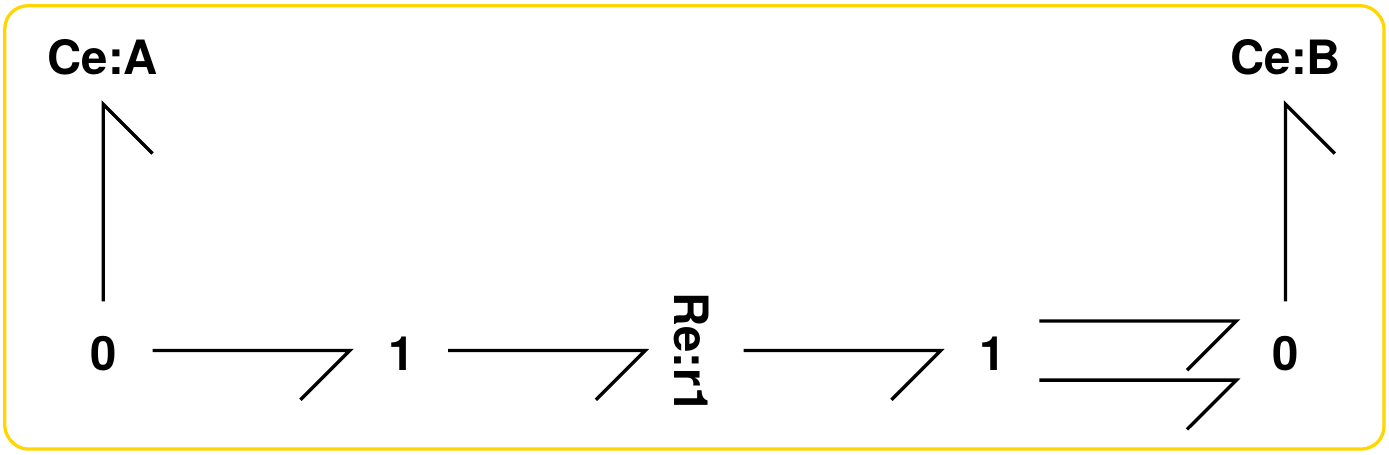}{\ch{A <> [ r1 ] 2 B} (Module M1)}{0.4}
  \SubFig{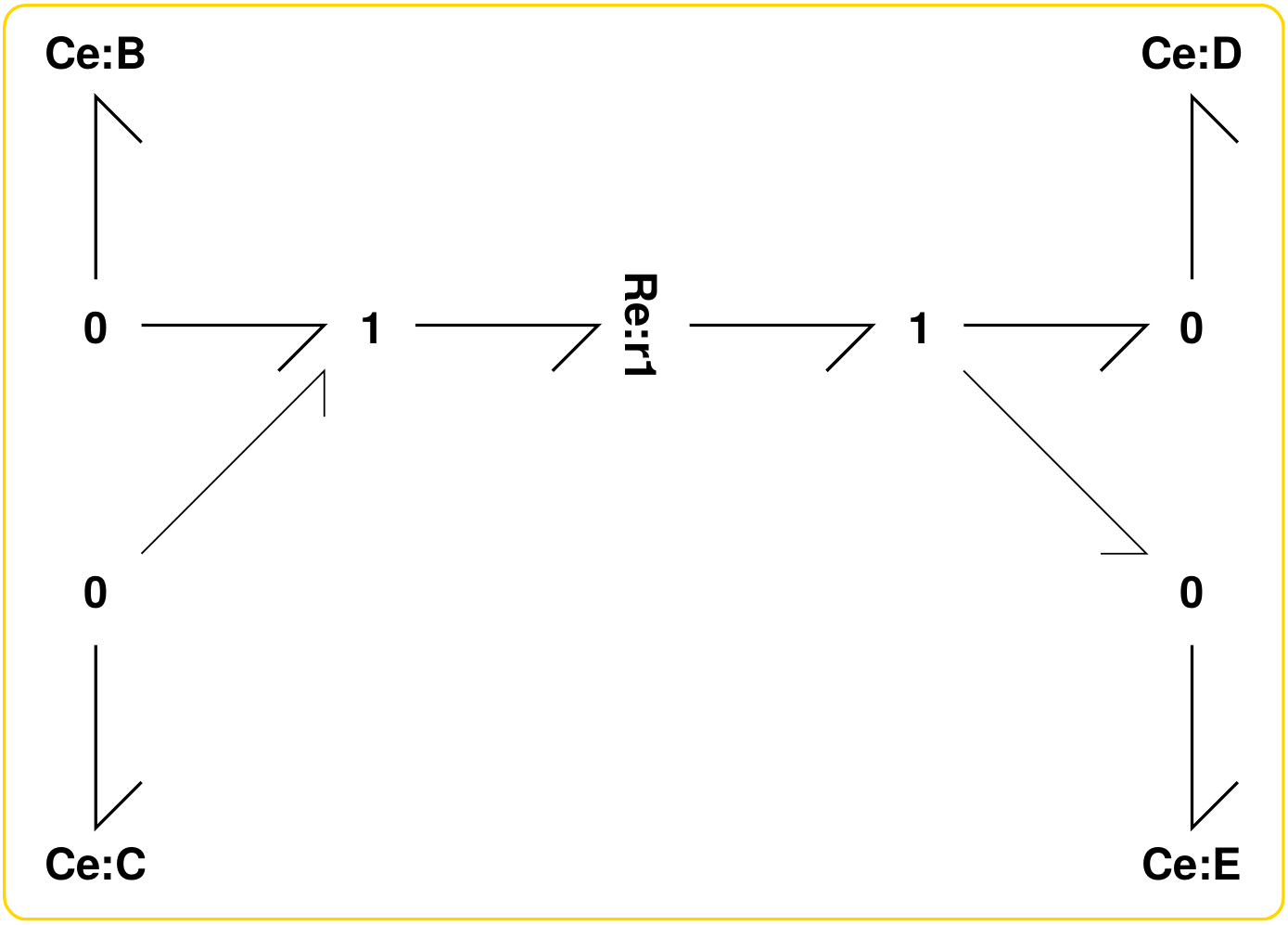}{\ch{B + C <> [ r2 ] D + E} (Module M2)}{0.4}
  \caption{Bond graphs of simple reactions.}
\end{figure}

In other words, the stoichiometric matrix $\NN$ can be derived from
the system bond graph.
This section shows that, conversely, the system bond graph can
be derived from the stoichiometric matrix $\NN$. 
The following
constructive procedure is used:
\begin{framed}
\begin{enumerate}
\item For each \emph{species} create a \Ce component with appropriate
  name and a \zero junction; connect a bond from the \zero junction to
  the \Ce component.
\item For each \emph{reaction} create an \Re component with appropriate
  name and two  \one junctions; connect a bond from one \one junction to
  the forward port of the \Re component and a bond from the reverse port of
  the \Re component to the other \one junction.
\item For each \emph{negative} entry $N_{ij}$ in the stoichiometric matrix,
  connect $-N_{ij}$ bonds from the zero junction connected to the
  $i$th species to the the one junction connected to the forward port
  of the $j$th reaction.
\item For each \emph{positive} entry $N_{ij}$ in the stoichiometric matrix,
  connect $N_{ij}$ bonds from the one junction connected to the reverse port
  of the $j$th reaction to the zero junction connected to the
  $i$th species.
\end{enumerate}
\end{framed}
For example, the reaction \ch{A <> [ r1 ] 2 B} has the stoichiometric
matrix
\begin{align}
N &=
    \left(
    \begin{matrix}
      -1\\
      2
    \end{matrix}
  \right)
\end{align}
and the bond graph of
Figure \ref{subfig:AB_abg}.
The reaction \ch{B + C <> [ r2 ] D + E} has the stoichiometric
matrix
\begin{align}
N &=
    \left(
    \begin{matrix}
      -1\\-1\\1\\1
    \end{matrix}
  \right)
\end{align}
and has the bond graph of Figure \ref{subfig:BCDE_abg}.

\section{Chemostats, Flowstats and Pathways}
\label{sec:chemostats-flowstats}
As discussed previously \citep{GawCra16,Gaw17a}, the notion of a \emph{chemostat}
\citep{PolEsp14} is useful in creating an open system from a closed
system.
As discussed by \citet{Gaw17a}, the chemostat has a number of
interpretations:
\begin{enumerate}
\item one or more species are fixed to give a constant concentration
  \citep{GawCurCra15}; this implies that an appropriate external
  flow is applied to balance the internal flow of the species.
% \item an ideal feedback controller is applied to species to be fixed
%   with setpoint as the fixed concentration and control signal an
%   external flow.
\item as a \Ce component with a fixed state.
% \item as an ideal source of chemical \potential.
\item as an external \emph{port} of a module which allows connection
  to other modules.
\end{enumerate}
In the context of stoichiometric analysis, the chemostat concept
provides a flexible alternative to the primary and currency exchange
reactions \citep{SchLetPal00,Pal06,Pal15}.

\citet{GawCra16} discuss the dual concept of \emph{flowstats} which
again  has a number of
interpretations:
\begin{enumerate}
\item one or more reaction flows are fixed.
% \item an ideal feedback controller is applied to reaction flows to be fixed
%   with setpoint as the fixed flow and control signal an
%   external \potential.
\item as an \Re component with a fixed flow.
%\item as an ideal source of reaction flow.
\item as an external \emph{port} of a module which allows connection
  to other modules.
\end{enumerate}
In the context of stoichiometric analysis, the flowstat concept
provides a way of isolating parts of a network by setting zero flow in
the reactions connecting the parts. Such zero flow flowstats can also
be interpreted as removing the corresponding enzyme via gene knockout.

In terms of stoichiometric analysis, the closed system equations
(\ref{eq:dX_closed}) and (\ref{eq:Phi_closed}) are replaced by:
\begin{align}
  \dot{X} &= \Ncd V \label{eq:dX}\\
  \Phi &= -N^T \phi\label{eq:Phi}
\end{align}
where $N^{cd}$ is created from the stoichiometric matrix $N$ by
setting \emph{rows} corresponding to chemostats species and
\emph{columns} corresponding to flowstatted reactions to zero
\citep{GawCra16}.
As discussed by \cite{GawCra16}, system pathways corresponding to
(\ref{eq:dX}) are defined by the right-null space of $N^{cd}$ that is
the columns of the matrix $K^{cd}$ where $N^{cd}K^{cd}=0$.
Further, then steady-state
pathways are defined by:
\begin{equation}\label{eq:V}
  V = K^{cd} v
\end{equation}
were $v$ is the pathway flow. It follows from Equation (\ref{eq:dX})
that Equation (\ref{eq:V}) implies that $\dot{X}=0$.
\citet{GawCra17} define the \emph{pathway} stoichiometric matrix $\Np$ as:
\begin{equation}\label{eq:N_p}
  N_p = N K^{cd}
\end{equation}
In a similar fashion to equation (\ref{eq:Phi}), the pathway reaction
\potentials $\Phi_p$ are given by
\begin{equation}
  \Phi_p = -N_p^T \phi\label{eq:Phip}
\end{equation}
In the same way as the stoichiometric matrix $\NN$ relates reaction flows to
species and thus represents a set of reactions, the pathway
stoichiometric matrix $\Np$ also represents a set of reactions: these
reactions will be called the \emph{pathway reactions}.

Following \citet{SchLetPal00}, pathways can be divided into three
categories according to the species corresponding to the non zero
elements in the relevant column of the \emph{pathway} stoichiometric matrix $\Np$:
\begin{description}
\item[I] The species include primary metabolites; these pathways are
  of functional interest.
\item[II] The species include currency metabolites only; these
  pathways dissipate energy without creating or consuming primary
  metabolites. \citet{SchLetPal00} call these pathways \emph{futile
    cycles}.
\item[III] There are no species.
\end{description}
Pathway reactions for type I pathways contain both primary and
currency metabolites; pathway reactions for type II pathways contain 
currency metabolites only; pathway reactions for type III pathways are empty.

Pathways have an equivalent bond graph obtained by applying the
conversion method of \S~\ref{sec:bond-graphs-integr} to $\Np$ instead
of $\NN$ \cite{GawCra17}; this fact can be utilised to give simple
physically plausible models of complex systems \cite{GawCudCra19X}.

\subsection{Illustrative example \cite{Noo18}}
\label{sec:ilustr-example-citen}
\begin{figure}[htbp]
  \centering
  \SubFig{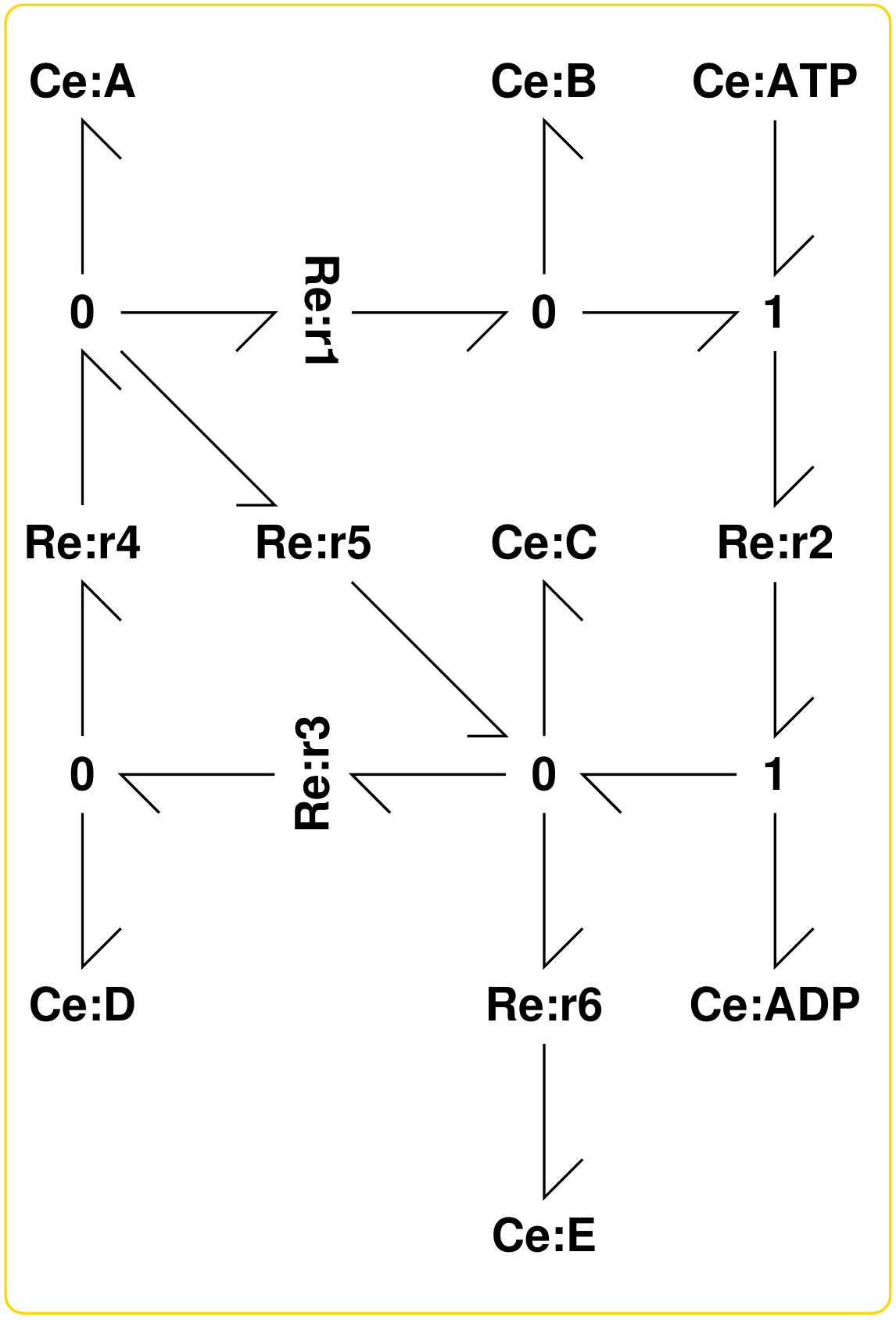}{Bond graph}{0.25}
  \SubFig{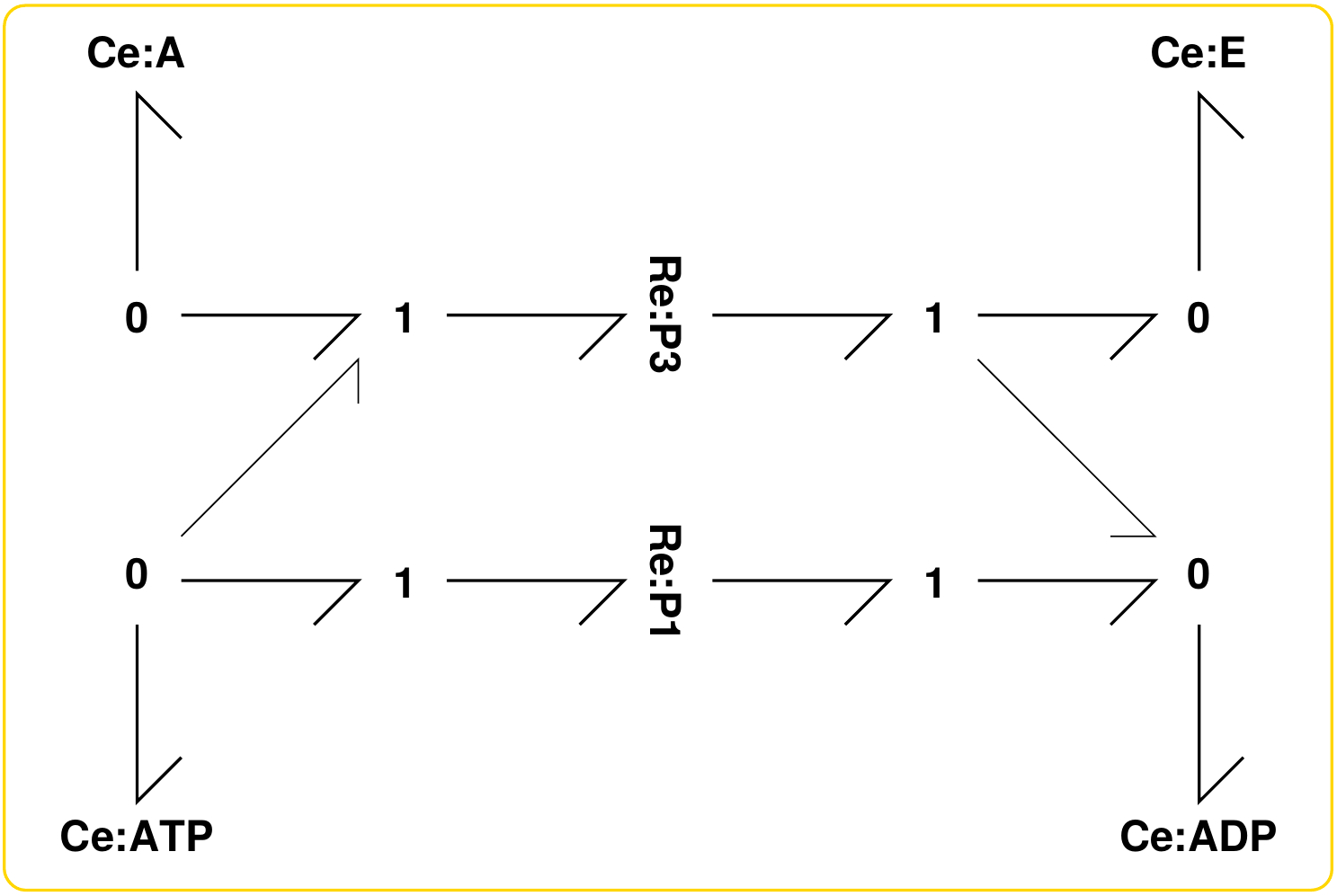}{Pathway bond graph}{0.35}
  \caption{Bond graphs for illustrative example \cite{Noo18}}
  \label{fig:Toy_abg}
\end{figure}
\citet{Noo18} gives a simple illustrative example of the three types
of pathway;
Figure \ref{subfig:Toy_abg} gives the corresponding bond graph.
the reactions are:
\begin{align}
\ch{A &<>[ r1 ] B }\\
\ch{ATP + B &<>[ r2 ] ADP + C }\\
\ch{C &<>[ r3 ] D }\\
\ch{D &<>[ r4 ] A }\\
\ch{A &<>[ r5 ] C }\\
\ch{C &<>[ r6 ] E }
\end{align}
The there are seven species and six reactions giving states $\xx$ and
flows $\vv$:
\begin{xalignat}{2}
  \xx&= \begin{pmatrix}
    x_{A}\\
    x_{ADP}\\
    x_{ATP}\\
    x_{B}\\
    x_{C}\\
    x_{D}\\
    x_{E}\\
  \end{pmatrix}&
  \vv&= \begin{pmatrix}
    v_{r1}\\
    v_{r2}\\
    v_{r3}\\
    v_{r4}\\
    v_{r5}\\
    v_{r6}\\
\end{pmatrix}
\end{xalignat}
The stoichiometric matrix is:
\begin{align}
N &=
    \left(\begin{matrix}
        -1 & 0 & 0 & 1 & -1 & 0\\
        0 & 1 & 0 & 0 & 0 & 0\\
        0 & -1 & 0 & 0 & 0 & 0\\
        1 & -1 & 0 & 0 & 0 & 0\\
        0 & 1 & -1 & 0 & 1 & -1\\
        0 & 0 & 1 & -1 & 0 & 0\\
        0 & 0 & 0 & 0 & 0 & 1
      \end{matrix}\right)
\end{align}
%
% The example is extended by assigning a set of nominal chemical
% \potentials $\phis$ to the species: $\phis_A=1$, $\phi_{ATP}=0$,
% $\phi_{ADP}=3$, $\phis_B=1$, $\phis_C=1$, $\phis_D=1$, $\phis_E=0$.
% %
% Using (\ref{eq:Phi}), the coresponding reaction \potential $\Phi$ can
% be computed
% \begin{xalignat*}{2}
% \ch{A &<>[ r1 ] B }&&(0.00)\\
% \ch{ATP + B &<>[ r2 ] ADP + C }&&(-2.00)\\
% \ch{C &<>[ r3 ] D }&&(0.00)\\
% \ch{D &<>[ r4 ] A }&&(0.00)\\
% \ch{A &<>[ r5 ] C }&&(0.00)\\
% \ch{C &<>[ r6 ] E }&&(1.00)
% \end{xalignat*}

Setting \ch{A}, \ch{E}, \ch{ATP} and \ch{ADP} as chemostats, $\Ncd$ is
constructed by setting the corresponding rows of $\NN$ to zero. The
corresponding null space is three dimensional and corresponds to the
three pathways:
\begin{enumerate}
\item r1 + r2 + r3 + r4
\item r3 + r4 + r5
\item r1 + r2 + r6
\end{enumerate}

Using (\ref{eq:N_p}), the pathway stoichiometric matrix $\Np$ is:
\begin{align}
\Np &=
    \left(\begin{matrix}0 & 0 & -1\\
        1 & 0 & 1\\-1 & 0 & -1\\
        0 & 0 & 0\\0 & 0 & 0\\
        0 & 0 & 0\\0 & 0 & 1\end{matrix}\right)
\end{align}

The three pathway reactions are:
\begin{align}
\ch{ATP &<>[ P1 ] ADP }\\
\ch{&<>[ P2 ] }\\
\ch{A + ATP &<>[ P3 ] ADP + E }
\end{align}
Pathway reaction P1 corresponds to a type II pathway, pathway reaction
P2 to a type III pathway and pathway reaction P3 to a type I pathway
where \ch{A} is converted to \ch{E} driven by the conversion of
\ch{ATP} to \ch{ADP}.
The example is extended by assigning a set of nominal chemical
\potentials $\phis$ to the species: $\phis_A=1$, $\phi_{ATP}=0$,
$\phi_{ADP}=3$, $\phis_B=1$, $\phis_C=1$, $\phis_D=1$, $\phis_E=0$.
The pathway reaction \potentials are then computed using
(\ref{eq:Phip}) as $\Phi_{P1}= -2$, $\Phi_{P2}= 0$, $\Phi_{P3}= -1$.
As the \potential for each pathway only depends on the species appearing
in the pathway reactions, the \potential of non-chemostatted species
are irrelevant for this computation. In fact the \potentials of the
species will correspond to the steady-state values of concentrations of
the non-chemostatted species arising from the flow patterns
corresponding to the chemostat \potentials \citep{Gaw18}.
The pathway bond graph  appears in Figure \ref{subfig:ToyPath_abg}.

\subsection{Example: Glycolysis \& Pentose Phosphate Pathways}
\label{sec:example:-glycolysis-}
The combination of the Glycolysis \& Pentose Phosphate networks
provides a number of different products from the metabolism of
glucose. This flexibility is adopted by proliferating cells, such as
those associated with cancer, to adapt to changing requirements of
biomass and energy production \citep{HeiCanTho09}.

The \emph{e.coli} Core Model \citep{OrtFlePal10,Pal15} is used as the
basis for the examples in this section. In particular, the species,
reactions and stoichiometric matrix were extracted from the
spreadsheet \texttt{ecoli\_core\_model.xlsx} but with the biomass
equations deleted and the the reaction \ch{CYTBD} (containing
$\frac{1}{2}\ch{O2}$) multiplied by 2 to give integer
stoichiometry. The submodel containing the reactions of the combined
Glycolysis \& Pentose Phosphate pathways was then extracted (see
Appendix~\ref{sec:glycolysis--pentose} for details) and converted to a
bond graph in bond graph tools format using the algorithm of
\S~\ref{sec:bond-graphs-integr}.
The following procedure was adopted to obtain
physiologically-realistic values for the species \potentials $\phi$.
\begin{enumerate}
\item The \emph{reaction} \potentials $\Phi$ were extracted from Table 4
  provided by \citet{ParRubXu16}.
\item A set of consistent \emph{species} \potentials $\phi$ was
  obtained from equation (\ref{eq:Phi}) using
  \begin{equation}
    \phi = - \lb N^T \rb^\dagger \Phi
  \end{equation}
  where $^\dagger$ denotes the pseudo inverse%
  \footnote{The pseudo inverse was implemented using the python linear
  algebra package function \texttt{linalg.pinv()}}.
\end{enumerate}

The reaction \potentials for each reaction are given in
Appendix~\ref{sec:glycolysis--pentose} and, because of the above
procedure,  correspond to the reaction \potentials listed by
\citet{ParRubXu16} Table 4.

As discussed by \citet[\S~22.6d]{GarGri17}, it illuminating to pick
out individual paths through the network to see how these may be
utilised to provide a variety of products. This is reproduced here by
choosing appropriate chemostats and flowstats
(\S~\ref{sec:chemostats-flowstats}) to give the results listed by
\citet[\S~22.6d]{GarGri17}. In each case, the corresponding pathway
reaction \potential is given. For consistency with
\citet[\S~22.6d]{GarGri17}, each pathway starts with Glucose
6-phosphate \Ch{G6P}.

The following chemostat list is used (together with additional
chemostats) in each of the following sections:
\{
\ch{ADP},
\ch{ATP},
\ch{CO2},
\ch{G6P},
\ch{H},
\ch{H2O},
\ch{NAD},
\ch{NADH},
\ch{NADP},
\ch{NADPH},
\ch{PI},
\ch{PYR}\}.

\subsubsection{Glycolysis}
The glycolysis pathway is isolated from the pentose phosphate pathway
by replacing the two connecting reactions (G6PDH2R and TKT2) by
flowstats.
\Pathway
{}
{G6PDH2R, TKT2}
{
\item PGI + PFK + FBA + TPI + 2GAPD - 2PGK - 2PGM + 2ENO + 2PYK
}
{
\ch{3 ADP + G6P + 2 NAD + 2 PI &<>[ P1 ] 3 ATP + H + 2 H2O + 2 NADH + 2 PYR }&&(-42.22~\si{\kilo\joule\per\mol})
}
The pathway reaction \ch{P_1} is the overall glycolysis reaction
\citet[\S~18.2]{GarGri17}. The negative reaction \potential indicates
that the reaction proceeds in the forward direction.

\subsubsection{\ch{R5P} \& \ch{NADPH} generation}
This pathway is isolated by setting PGI and TKT2 as flowstats and the
product \ch{R5P} is added to the chemostat list.
\Pathway
{\ch{R5P}}
{PGI, TKT2}
{
\item G6PDH2R + PGL + GND + RPI
}
{
\ch{G6P + H2O + 2 NADP &<>[ P1 ] CO2 + 2 H + 2 NADPH + R5P }&&(-17.01~\si{\kilo\joule\per\mol})
}
The pathway reaction \ch{P_1} corresponds to the \ch{R5P} \&
\ch{NADPH} synthesis discussed in comment 1 of
\citet[\S~22.6d]{GarGri17}.
The negative reaction \potential indicates
that the reaction proceeds in the forward direction.

\subsubsection{\ch{R5P} generation}
This pathway is isolated by setting GAPD and G6PDH2R as flowstats and the
product \ch{R5P} is added to the chemostat list.
\Pathway
{\ch{G6P}, \ch{R5P}}
{GAPD, G6PDH2R}
{
\item  - 5PGI - PFK - FBA - TPI - 4RPI + 2TKT2 + 2TALA + 2TKT1 + 4RPE
}
{
  \ch{ADP + H + 6 R5P &<>[ P1 ] ATP + 5 G6P }&&(20.30~\si{\kilo\joule\per\mol})
}
The pathway reaction \ch{P_1} corresponds to the \ch{R5P} synthesis discussed in comment 2 of
\citet[\S~22.6d]{GarGri17}.
The positive reaction \potential indicates
that the reaction proceeds in the reverse direction.

\subsubsection{\ch{NADPH} generation}
This pathway is isolated by setting GAPD as a flowstat.
\Pathway
{}
{GAPD}
{
\item - 5PGI - PFK - FBA - TPI + 6G6PDH2R + 6PGL + 6GND + 2RPI + 2TKT2 + 2TALA + 2TKT1 + 4RPE
}
{
 \ch{ADP + G6P + 6 H2O + 12 NADP &<>[ P1 ] ATP + 6 CO2 + 11 H + 12 NADPH }&&(-81.79~\si{\kilo\joule\per\mol}) 
}
The pathway reaction \ch{P_1} corresponds to the 
\ch{NADPH} synthesis discussed in comment 3 of
\citet[\S~22.6d]{GarGri17}.
The negative reaction \potential indicates
that the reaction proceeds in the forward direction.

\subsubsection{\ch{NADPH} \& \ch{ATP} generation}
This pathway is isolated by setting PGI as flowstat.
\Pathway
{}
{PGI}
{
\item 2PFK + 2FBA + 2TPI + 5GAPD - 5PGK - 5PGM + 5ENO + 5PYK + 3G6PDH2R + 3PGL + 3GND + RPI + TKT2 + TALA + TKT1 + 2RPE
}
{
  &\ch{8 ADP + 3 G6P + 5 NAD + 6 NADP + 5 PI <>[ P1 ] } && \notag\\
  &\ch{8 ATP + 3 CO2 + 8 H + 2 H2O + 5 NADH + 6 NADPH + 5 PYR }&&(-146.44~\si{\kilo\joule\per\mol})
}
The pathway reaction \ch{P_1} corresponds to the 
\ch{NADPH} and \ch{ATP} synthesis discussed in comment 4 of
\citet[\S~22.6d]{GarGri17}.
The negative reaction \potential indicates
that the reaction proceeds in the forward direction.

\section{Modularity}
\label{sec:modularity}
As discussed by \citet{GawCra16}, there are two related but
distinct concepts of modularity: computational modularity where
physical correctness is retained and behavioural modularity where
module behaviour (such as ultra-sensitivity) is retained. It is the
former that is discussed in this section.
As discussed by \citet{Gaw17a}, modular bond graphs provide a way of
decomposing complex biomolecular systems into manageable
parts~\citep{GawCurCra15,GawCra16}. In particular, this paper combines
the modularity concepts of \citet{NeaCarTho16} with the bond graph
approach to give a more flexible approach to modularity.
The basic idea \citep{Gaw17a} is simple: modules are self-contained
and have no explicit ports; but any species, as represented by a \Ce
component has the potential to become a port. Thus if two modules
share the same species, the corresponding \Ce component in each module
is replaced by a port with the same name, and the species is
explicitly represented as a \Ce component on a higher level. Moreover,
each module can be individually tested by replacing the relevant \Ce
components by chemostats. 

The algorithm is:
\begin{enumerate}
\item Within each module, each \Ce component corresponding to a common
  species is \emph{exposed} -- replaced by a \emph{port}
  component. Note that the algorithm of \S~\ref{sec:bond-graphs-integr}
  ensures that each \Ce is attached to a \zero junction.
\item For each common species, create a \Ce component connected to a
  \zero component.
\item Connect all module ports associate with each species to the \zero
  junction associated with the  species; all instances of \Ce
  components corresponding to each species are thus \emph{unified}.
\end{enumerate}
\begin{figure}[htbp]
  \centering
  \Fig{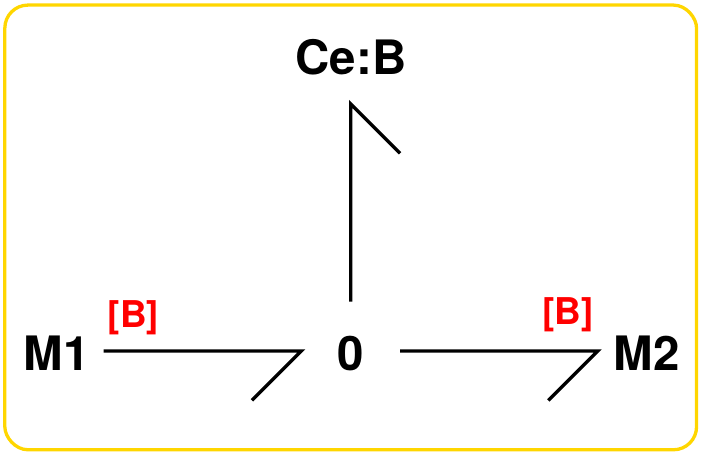}{0.3}
  \caption{Modularity. Modules M1 and M2 correspond to Figures
    \ref{subfig:AB_abg} \& \ref{subfig:BCDE_abg} respectively. The
  common species \ch{B} is exposed as a port in each module and
  connected to the new \BCe{B} component via a \zero junction.}
  \label{fig:modularity}
\end{figure}
For example, let modules M1 and M2 correspond to Figures
\ref{subfig:AB_abg} \& \ref{subfig:BCDE_abg} respectively. In Figure
\ref{fig:modularity}, the common
species \ch{B} is exposed as a port in each module and connected to
the new \BCe{B} component via a \zero junction. The composite system
contains the two reactions:
\begin{align}
\ch{A &<>[ r1 ] 2 B }\\
\ch{C + B &<>[ r1 ] D + E }
\end{align}
Choosing the set of chemostats to be
$\left\{ \ch{A}, \ch{C}, \ch{D}, \ch{E}\right\}$
the corresponding pathway stoichiometric matrix $\Np$ is
\begin{align}
\Np &=
\left(\begin{matrix}-1\\-2\\2\\2\\0\end{matrix}\right)
\end{align}
where the species are
$\left\{ \ch{A}, \ch{C}, \ch{D}, \ch{E}, \ch{B}\right\}$ and the
reactions $\left\{ \ch{r1}, \ch{r2}\right\}$. The pathway reaction
\ch{P1} is then:
\begin{align}
\ch{A + 2 C &<>[ P1 ] 2 D + 2 E }
\end{align}

\subsection{ Example: Metabolism}\label{sec:example:-metabolism}

As in \S~\ref{sec:example:-glycolysis-}, the \emph{e.coli} Core Model
\citep{OrtFlePal10,Pal15} is used. In particular, reactions
corresponding to four modules (Glycolysis, TCA cycle, Electron
Transport Chain and ATPase) were extracted as detailed in Appendix
\ref{sec:modul-repr-metab}.
For simplicity, reaction PDH (converting
\ch{PYR} to \ch{ACCOA}) and reaction NADTRHD (converting
\ch{NADP}/\ch{NADPH}n to \ch{NAD}/\ch{NADH}) were included in the TCA
  cycle module.

These modules can be analysed individually. For example the TCA cycle
module can be analysed using the set of chemostats:
\begin{equation*}
  \{\ch{PYR}, \ch{CO2}, \ch{ADP}, \ch{ATP}, \ch{H2O},
\ch{NAD}, \ch{NADH}, \ch{PI}, \ch{H},
\ch{Q8},\ch{Q8H2} \}
\end{equation*}
The two pathways are
\begin{enumerate}
\item  FRD7 + SUCDI
\item PDH + CS + ACONTA + ACONTB + ICDHYR + AKGDH - SUCOAS - FRD7 + FUM + MDH + NADTRHD
\end{enumerate}
These two pathways correspond to the two pathway reactions:
\begin{align*}
\ch{&<>[ P1 ] }\\
  % \ch{&ADP + 2 H2O + 4 NAD + PI + PYR + Q8 <>[ P2 ]}\\
  % \ch{&ATP + 3 CO2 + 2 H + 4 NADH + Q8H2}
\ch{ADP + 2 H2O + 4 NAD + PI + PYR + Q8 &<>[ P2 ] ATP + 3 CO2 + 2 H + 4 NADH + Q8H2 }p
\end{align*}
The first is a type III reaction and the second a type I reaction
which utilises the \potential of \ch{PYR} to generate two \ch{NADH}, one
\ch{NADHP}, one {ATP} and one \ch{Q8H2} whilst releasing two \ch{CO2} and
two \ch{H}.

The overall metabolic system comprises the four modules (Glycolysis, TCA cycle, Electron
Transport Chain and ATPase) connected together. Using the approach of
\S~\ref{sec:modularity}, the modules are interconnected by declaring
the set of species that the modules have in common:
\begin{equation*}
  \{\ch{PYR},\ch{ATP},\ch{ADP},\ch{PI},\ch{H},\ch{H_E},\ch{NAD},\ch{NADH},\ch{H2O},\ch{Q8},\ch{Q8H2}\}
\end{equation*}
These species are unified as described in \S~\ref{sec:modularity}.
To analyse the composite system, the set of chemostats was chosen as:
\begin{equation*}
  \{\ch{GLCD_E},\ch{CO2},\ch{O2},\ch{ADP},\ch{ATP},\ch{H2O},\ch{PI},\ch{H}\}.
\end{equation*}
The three pathways are
\begin{enumerate}
\item PFK + FBP
\item  FRD7 + SUCDI
\item 2 GLCPTS + 2 PGI + 2 PFK + 2 FBA + 2 TPI + 4 GAPD - 4 PGK - 4 PGM + 4 ENO + 2 PYK + 4 PDH + 4 CS + 4 ACONTA + 4 ACONTB + 4 ICDHYR + 4 AKGDH - 4 SUCOAS - 4 FRD7 + 4 FUM + 4 MDH + 4 NADTRHD + 20 NADH16 + 12 CYTBD + 27 ATPS4R
\end{enumerate}
These three pathways correspond to the three pathway reactions:
\begin{align*}
\ch{ATP + H2O &<>[ P1 ] ADP + PI + H }\\
\ch{&<>[ P2 ] }\\
\ch{2 GLCD_E + 12 O2 + 35 ADP + 35 PI + 35 H &<>[ P3 ] 12 CO2 + 35 ATP + 47 H2O }
\end{align*}
As in \S~\ref{sec:ilustr-example-citen}, pathway reaction P1
corresponds to a type II pathway, pathway reaction P2 to a type III
pathway and pathway reaction P3 to a type I pathway.
Pathway 3 corresponds to the metabolic generation of \ch{ATP} using the
free energy of \ch{GLCD_E}.
The ratio of \ch{ATP} to \ch{GLCD_E} is 17.5; this is the value quoted
by \citet[\S~19.2]{Pal15}.

\section{FBA and EBA in a bond graph  context}
\label{sec:eba-bond-graph}
%\subsection{FBA and EBA}
The standard FBA approach is to create open systems from closed
systems by adding ``exchange reactions
'' to species which connect to
the outside world -- for example: \ch{ATP <> $\oslash$}. In contrast, the bond
graph approach would declare \ch{ATP} to be a chemostat.
Chemostats provide a more flexible approach as they can be created
without changing system structure and are used in the sequel.

FBA \citep{OrtThiPal10} uses the linear equation (\ref{eq:V}) within a constrained
linear optimisation to compute pathway flows. EBA adds two sorts of
nonlinear constraint arising from thermodynamics.
This section shows that the bond graph  approach automatically
includes the EBA constraint equations by considering Inequality
(\ref{eq:diss}) and  Equation
  (\ref{eq:Phi}). In particular:
\begin{enumerate}
\item Inequality (\ref{eq:diss}) corresponds to Equation 8 of
  \cite{BeaLiaQia02}.
  This inequality can be re-expressed as:
  \begin{align}
    \Phi_i  &= r_i(\phi) V_i \\
    \where r_i(\phi)& >0\label{eq:r>0}
  \end{align}
  $r_i$ corresponds to the ``flux resistances'' on p.83 of
  \cite{BeaLiaQia02}].
\item If $K$ is the right null matrix of $N$, it follows from Equation
  (\ref{eq:Phi}) that
  \begin{equation}
    K^T \Phi = 0
  \end{equation}
  This corresponds to Equation 7 of \citet{BeaLiaQia02}.
  Note that $K$ defines the pathways of the closed system system (with no chemostats).
\end{enumerate}

Moreover, the pathways of the open system as defined by $\Kcd$ can be
considered by defining $R = \diag~{r_i}$ and using Equation
(\ref{eq:V}):
\begin{equation}\label{eq:constraint}
  \boxed{K^T R K^{cd} v = 0}
\end{equation}
Equation (\ref{eq:constraint}) and inequality (\ref{eq:r>0}) constrain
the pathway flows $v$; this is illustrated in the following examples
drawn from \citet{BeaLiaQia02}.

\subsection{Example: Parallel reactions}
\label{sec:exampl-parall-react}
\begin{figure}[htbp]
  \centering
  \SubFig{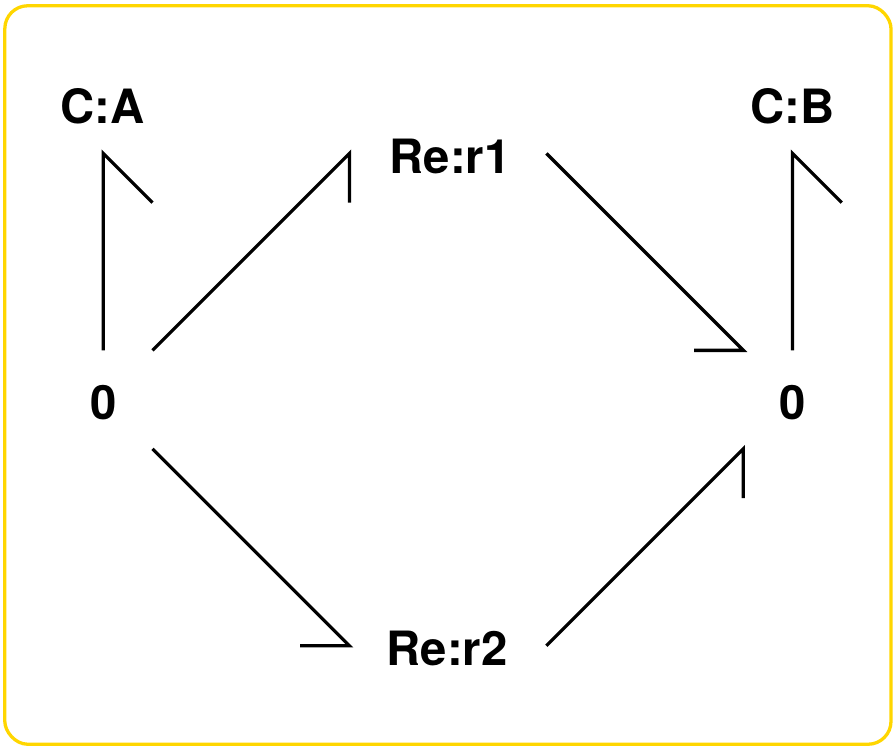}{Example: Parallel reaction.}{0.25}
  \SubFig{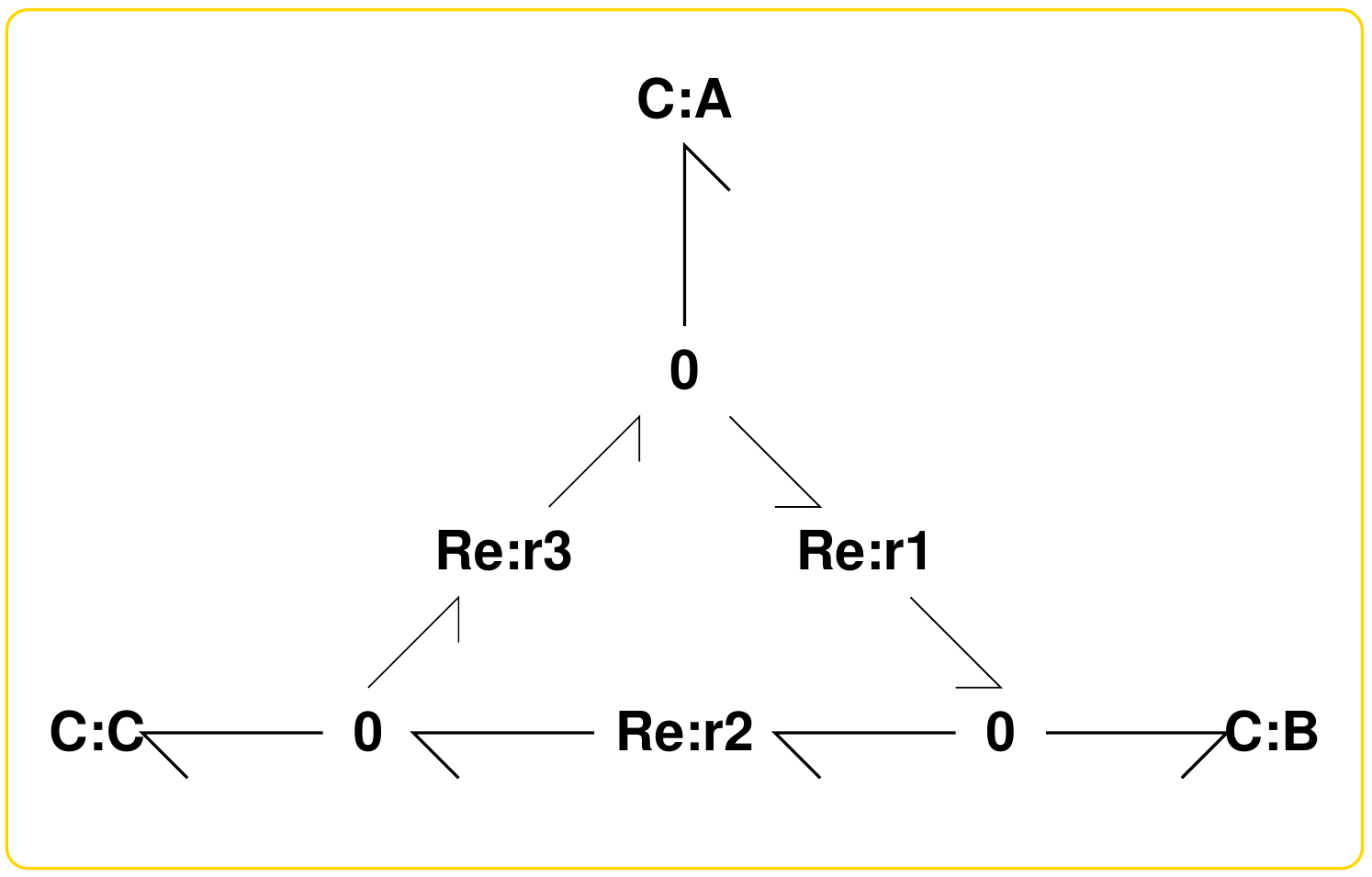}{Example: three-reaction cycle.}{0.35}
  \caption{Bond graphs coresponding to examples from
    \citet{BeaLiaQia02} (\one junctions are not shown for clarity)).
    (a) \citep[Fig. 2]{BeaLiaQia02},
    (b) \citep[Fig. 3]{BeaLiaQia02}
  }
  \label{fig:pAB}
\end{figure}
\citet[Fig. 2]{BeaLiaQia02} motivate EBA using the example of two
resistors in parallel. Figure \ref{subfig:pAB_abg} shows the bond graph  of
the analogous reaction system: the species \ch{A} and \ch{B} are
joined by two reactions:
\begin{align}
\ch{A &<>[ r1 ] B }\\
\ch{A &<>[ r2 ] B }
\end{align}
The stoichiometric matrix is:
\begin{align}
N &=
\left(\begin{matrix}-1 & -1\\1 & 1\end{matrix}\right)
\end{align}
and the null space matrix $K$ is
\begin{align}
K &=
\left(\begin{matrix}-1\\1\end{matrix}\right)
\end{align}
corresponding to the pathway: $-r_1 + r_2$.

Setting \ch{A} and \ch{B} as chemostats:
\begin{align}
N^{cd} &=
\left(\begin{matrix}0 & 0\\0 & 0\end{matrix}\right)
\end{align}
\begin{align}
K^{cd} &=
\left(\begin{matrix}1 & 0\\0 & 1\end{matrix}\right)
\end{align}
Equation (\ref{eq:constraint}) then becomes:
\begin{equation}
  -r_1v_1 + r_2 v_2 = 0
\end{equation}
As $r_i>0$, it follows that $v_1$ and $v_2$ must either be zero or
have the same sign.

\subsection{Example: three-reaction cycle}
\label{sec:exampl-three-react}

\citet[Fig. 3]{BeaLiaQia02} give the example of a three-reaction
cycle. Figure \ref{subfig:ABC_abg} shows the corresponding bond graph.
The species \ch{A}, \ch{B} and \ch{C} are joined by three
reactions:
\begin{align}
\ch{A &<>[ r1 ] B }\\
\ch{B &<>[ r2 ] C }\\
\ch{C &<>[ r3 ] A }
\end{align}
The stoichiometric matrix is:
\begin{align}
N &=
\left(\begin{matrix}-1 & 0 & 1\\1 & -1 & 0\\0 & 1 & -1\end{matrix}\right)
\end{align}
and the null space matrix $K$ is
\begin{align}
K &=
\left(\begin{matrix}1\\1\\1\end{matrix}\right)
\end{align}
corresponding to the pathway: $r_1 + r_2 + r_3$.

Setting \ch{A} and \ch{B} as chemostats:
\begin{align}
N^{cd} &=
\left(\begin{matrix}0 & 0 & 0\\0 & 0 & 0\\0 & 1 & -1\end{matrix}\right)
\end{align}

\begin{align}
K^{cd} &=
\left(\begin{matrix}1 & 0\\0 & 1\\0 & 1\end{matrix}\right)
\end{align}
Equation (\ref{eq:constraint}) then becomes:
\begin{equation}
  r_1 v_1 + r_2 v_2 + r_3 v_2 = r_1 v_1 + \lb r_2 + r_3 \rb v_2 = 0
\end{equation}
As $r_i>0$, it follows that $v_1$ and $v_2$ must either be zero or
have the opposite sign.

Alternatively, setting \ch{A}, \ch{B} and \ch{C} as chemostats:
\begin{align}
N^{cd} &=
\left(\begin{matrix}0 & 0 & 0\\0 & 0 & 0\\0 & 0 & 0\end{matrix}\right)
\end{align}

\begin{align}
K^{cd} &=
\left(\begin{matrix}1 & 0 & 0\\0 & 1 & 0\\0 & 0 & 1\end{matrix}\right)
\end{align}

Equation (\ref{eq:constraint}) then becomes:
\begin{equation}
  r_1 v_1 + r_2 v_2 + r_3 v_3 = 0
\end{equation}
As $r_i>0$, there are three possibilities: all flows are zero; one of
the three pathway flows must have one sign and the other two flows the
opposite sign; or one flow is zero and the other two have opposite
signs.

\section{Conclusion}
\label{sec:conclusion}
\begin{enumerate}
\item It has been shown that the bond graph of a biomolecular system
  can be derived from the stoichiometric matrix. Thus the plethora of
  existing stoichiometric models can be automatically endowed with a
  number of features including
  \begin{enumerate}
  \item thermodynamic compliance
  \item modularity
  \item explicit energy flows allowing exploration of, for example,
    efficiency \citep{GawCra18}
  \item generation of reduced-order models using pathway analysis
    \citep{GawCra17,GawCudCra19X}.
  \item energy compliant connections to other physical domains
    including models of chemoelectric transduction
    \citep{GawSieKam17,Gaw17a}, membrane transporters
    \citep{PanGawTra19}, cardiac action potential \citep{PanGawTra18},
    chemomechanical transduction and photosynthesis.
  \end{enumerate}
\item The key equations of the  EBA approach of \citet{BeaLiaQia02}
  have been shown to be implicit in the system bond graph.
\item Via the modular approach of \S~\ref{sec:modularity}, the \Re
  components of \S~\ref{sec:bond-graphs-integr}, representing
  mass-action kinetics, can be replaced by thermodynamically compliant
  models of more complex kinetics \cite{Cor12} driven by enzymes and inhibitors
  including feedback inhibition, allosteric modulation and
  cooperativity.
\item This approach provides a basis for \emph{thermodynamically
    compliant} whole-cell models.
\end{enumerate}

  \section{Acknowledgements}
  I would like to thank the Melbourne School of
  Engineering for its support via a Professorial Fellowship, and
  Edmund Crampin and Michael Pan for help, advice and encouragement.

\bibliography{common}

\begin{thebibliography}{64}
\providecommand{\natexlab}[1]{#1}
\providecommand{\url}[1]{\texttt{#1}}
\expandafter\ifx\csname urlstyle\endcsname\relax
  \providecommand{\doi}[1]{doi: #1}\else
  \providecommand{\doi}{doi: \begingroup \urlstyle{rm}\Url}\fi

\bibitem[Atkins and de~Paula(2011)]{AtkPau11}
Peter Atkins and Julio de~Paula.
\newblock \emph{{Physical Chemistry for the Life Sciences}}.
\newblock Oxford University Press, 2nd edition, 2011.

\bibitem[Beard et~al.(2002)Beard, Liang, and Qian]{BeaLiaQia02}
Daniel~A. Beard, Shoudan Liang, and Hong Qian.
\newblock Energy balance for analysis of complex metabolic networks.
\newblock \emph{Biophysical Journal}, 83\penalty0 (1):\penalty0 79 -- 86, 2002.
\newblock ISSN 0006-3495.
\newblock \doi{10.1016/S0006-3495(02)75150-3}.

\bibitem[Borutzky(2010)]{Bor10}
Wolfgang Borutzky.
\newblock \emph{Bond graph methodology: development and analysis of
  multidisciplinary dynamic system models}.
\newblock Springer, Berlin, 2010.
\newblock ISBN 978-1-84882-881-0.
\newblock \doi{10.1007/978-1-84882-882-7}.

\bibitem[Cellier(1991)]{Cel91}
F.~E. Cellier.
\newblock \emph{Continuous system modelling}.
\newblock Springer-Verlag, New York, 1991.

\bibitem[Cornish-Bowden(2013)]{Cor12}
Athel Cornish-Bowden.
\newblock \emph{Fundamentals of enzyme kinetics}.
\newblock Wiley-Blackwell, London, 4th edition, 2013.
\newblock ISBN 978-3-527-33074-4.

\bibitem[Covert(2015)]{Cov15}
Markus~W. Covert.
\newblock \emph{Fundamentals of Systems Biology From Synthetic Circuits to
  Whole-cell Models}.
\newblock CRC Press, 2015.
\newblock \doi{10.4324/9781315222615}.

\bibitem[Cudmore et~al.(2019)Cudmore, Gawthrop, Pan, and
  Crampin]{CudGawPanCra19X}
Peter Cudmore, Peter~J. Gawthrop, Michael Pan, and Edmund~J. Crampin.
\newblock Computer-aided modelling of complex physical systems with
  {BondGraphTools}.
\newblock Available at arXiv:1906.10799, Jun 2019.

\bibitem[Dai and Locasale(2018)]{DaiLoc18}
Ziwei Dai and Jason~W. Locasale.
\newblock Thermodynamic constraints on the regulation of metabolic fluxes.
\newblock \emph{Journal of Biological Chemistry}, 293\penalty0 (51):\penalty0
  19725--19739, 2018.
\newblock \doi{10.1074/jbc.RA118.004372}.

\bibitem[Ebrahim et~al.(2013)Ebrahim, Lerman, Palsson, and Hyduke]{EbrLerPal11}
Ali Ebrahim, Joshua~A. Lerman, Bernhard~O. Palsson, and Daniel~R. Hyduke.
\newblock Cobrapy: Constraints-based reconstruction and analysis for python.
\newblock \emph{BMC Systems Biology}, 7\penalty0 (1):\penalty0 74, Aug 2013.
\newblock ISSN 1752-0509.
\newblock \doi{10.1186/1752-0509-7-74}.

\bibitem[Garrett and Grisham(2017)]{GarGri17}
Reginald~H. Garrett and Charles~M. Grisham.
\newblock \emph{Biochemistry}.
\newblock Cengage Learning, Boston, MA, 6th edition, 2017.

\bibitem[Gawthrop(2018)]{Gaw18}
P.~Gawthrop.
\newblock Computing biomolecular system steady-states.
\newblock \emph{IEEE Transactions on NanoBioscience}, 17\penalty0 (1):\penalty0
  36--43, March 2018.
\newblock ISSN 1536-1241.
\newblock \doi{10.1109/TNB.2017.2787486}.
\newblock Published online 25th December 2017.

\bibitem[Gawthrop(2017)]{Gaw17a}
P.~J. Gawthrop.
\newblock Bond graph modeling of chemiosmotic biomolecular energy transduction.
\newblock \emph{IEEE Transactions on NanoBioscience}, 16\penalty0 (3):\penalty0
  177--188, April 2017.
\newblock ISSN 1536-1241.
\newblock \doi{10.1109/TNB.2017.2674683}.
\newblock Available at {arXiv:1611.04264}.

\bibitem[Gawthrop and Crampin(2016)]{GawCra16}
P.~J. Gawthrop and E.~J. Crampin.
\newblock Modular bond-graph modelling and analysis of biomolecular systems.
\newblock \emph{IET Systems Biology}, 10\penalty0 (5):\penalty0 187--201,
  October 2016.
\newblock ISSN 1751-8849.
\newblock \doi{10.1049/iet-syb.2015.0083}.
\newblock Available at {arXiv:1511.06482}.

\bibitem[Gawthrop and Smith(1996)]{GawSmi96}
P.~J. Gawthrop and L.~P.~S. Smith.
\newblock \emph{Metamodelling: Bond Graphs and Dynamic Systems}.
\newblock Prentice Hall, Hemel Hempstead, Herts, England., 1996.
\newblock ISBN 0-13-489824-9.

\bibitem[Gawthrop et~al.(2017)Gawthrop, Siekmann, Kameneva, Saha, Ibbotson, and
  Crampin]{GawSieKam17}
P.~J. Gawthrop, I.~Siekmann, T.~Kameneva, S.~Saha, M.~R. Ibbotson, and E.~J.
  Crampin.
\newblock Bond graph modelling of chemoelectrical energy transduction.
\newblock \emph{IET Systems Biology}, 11\penalty0 (5):\penalty0 127--138, 2017.
\newblock ISSN 1751-8849.
\newblock \doi{10.1049/iet-syb.2017.0006}.
\newblock Available at {arXiv:1512.00956}.

\bibitem[Gawthrop and Bevan(2007)]{GawBev07}
Peter~J Gawthrop and Geraint~P Bevan.
\newblock Bond-graph modeling: A tutorial introduction for control engineers.
\newblock \emph{IEEE Control Systems Magazine}, 27\penalty0 (2):\penalty0
  24--45, April 2007.
\newblock \doi{10.1109/MCS.2007.338279}.

\bibitem[Gawthrop and Crampin(2014)]{GawCra14}
Peter~J. Gawthrop and Edmund~J. Crampin.
\newblock Energy-based analysis of biochemical cycles using bond graphs.
\newblock \emph{Proceedings of the Royal Society A: Mathematical, Physical and
  Engineering Science}, 470\penalty0 (2171):\penalty0 1--25, 2014.
\newblock \doi{10.1098/rspa.2014.0459}.
\newblock Available at {arXiv:1406.2447}.

\bibitem[Gawthrop and Crampin(2017)]{GawCra17}
Peter~J. Gawthrop and Edmund~J. Crampin.
\newblock Energy-based analysis of biomolecular pathways.
\newblock \emph{Proceedings of the Royal Society of London A: Mathematical,
  Physical and Engineering Sciences}, 473\penalty0 (2202), 2017.
\newblock ISSN 1364-5021.
\newblock \doi{10.1098/rspa.2016.0825}.
\newblock Available at {arXiv:1611.02332}.

\bibitem[Gawthrop and Crampin(2018)]{GawCra18}
Peter~J. Gawthrop and Edmund~J. Crampin.
\newblock Biomolecular system energetics.
\newblock In \emph{Proceedings of the 13th International Conference on Bond
  Graph Modeling ({ICBGM'18})}, Bordeaux, 2018. Society for Computer
  Simulation.
\newblock Available at {arXiv:1803.09231}.

\bibitem[Gawthrop et~al.(2015)Gawthrop, Cursons, and Crampin]{GawCurCra15}
Peter~J. Gawthrop, Joseph Cursons, and Edmund~J. Crampin.
\newblock Hierarchical bond graph modelling of biochemical networks.
\newblock \emph{Proceedings of the Royal Society A: Mathematical, Physical and
  Engineering Sciences}, 471\penalty0 (2184):\penalty0 1--23, 2015.
\newblock ISSN 1364-5021.
\newblock \doi{10.1098/rspa.2015.0642}.
\newblock Available at {arXiv:1503.01814}.

\bibitem[{Gawthrop} et~al.(2019){Gawthrop}, {Cudmore}, and
  {Crampin}]{GawCudCra19X}
Peter~J. {Gawthrop}, Peter {Cudmore}, and Edmund~J. {Crampin}.
\newblock {Physically-Plausible Modelling of Biomolecular Systems: A
  Simplified, Energy-Based Model of the Mitochondrial Electron Transport
  Chain}.
\newblock May 2019.

\bibitem[Goldberg et~al.(2018)Goldberg, Szigeti, Chew, Sekar, Roth, and
  Karr]{GolSziChe18}
Arthur~P Goldberg, Balázs Szigeti, Yin~Hoon Chew, John~AP Sekar, Yosef~D Roth,
  and Jonathan~R Karr.
\newblock Emerging whole-cell modeling principles and methods.
\newblock \emph{Current Opinion in Biotechnology}, 51:\penalty0 97 -- 102,
  2018.
\newblock ISSN 0958-1669.
\newblock \doi{10.1016/j.copbio.2017.12.013}.
\newblock Systems biology, Nanobiotechnology.

\bibitem[Heinrich and Schuster(1996)]{HeiSch96}
Reinhart Heinrich and Stefan Schuster.
\newblock \emph{The regulation of cellular systems}.
\newblock Chapman \& Hall New York, 1996.

\bibitem[Heirendt et~al.(2019)Heirendt, Arreckx, Pfau, Mendoza, Richelle,
  Heinken, Haraldsdottir, Wachowiak, Keating, Vlasov, Magnusdottir, Ng,
  Preciat, {\AA}{\oe}agare, Chan, Aurich, Clancy, Modamio, Sauls, Noronha,
  Bordbar, Cousins, El~Assal, Valcarcel, Apaolaza, Ghaderi, Ahookhosh,
  Ben~Guebila, Kostromins, Sompairac, Le, Ma, Sun, Wang, Yurkovich, Oliveira,
  Vuong, El~Assal, Kuperstein, Zinovyev, Hinton, Bryant, Aragon~Artacho,
  Planes, Stalidzans, Maass, Vempala, Hucka, Saunders, Maranas, Lewis, Sauter,
  Palsson, Thiele, and Fleming]{HeiArrPfa19}
Laurent Heirendt, Sylvain Arreckx, Thomas Pfau, Sebastian~N. Mendoza, Anne
  Richelle, Almut Heinken, Hulda~S. Haraldsdottir, Jacek Wachowiak, Sarah~M.
  Keating, Vanja Vlasov, Stefania Magnusdottir, Chiam~Yu Ng, German Preciat,
  Alise {\AA}{\oe}agare, Siu H.~J. Chan, Maike~K. Aurich, Catherine~M. Clancy,
  Jennifer Modamio, John~T. Sauls, Alberto Noronha, Aarash Bordbar, Benjamin
  Cousins, Diana~C. El~Assal, Luis~V. Valcarcel, Ioigo Apaolaza, Susan Ghaderi,
  Masoud Ahookhosh, Marouen Ben~Guebila, Andrejs Kostromins, Nicolas Sompairac,
  Hoai~M. Le, Ding Ma, Yuekai Sun, Lin Wang, James~T. Yurkovich, Miguel A.~P.
  Oliveira, Phan~T. Vuong, Lemmer~P. El~Assal, Inna Kuperstein, Andrei
  Zinovyev, H.~Scott Hinton, William~A. Bryant, Francisco~J. Aragon~Artacho,
  Francisco~J. Planes, Egils Stalidzans, Alejandro Maass, Santosh Vempala,
  Michael Hucka, Michael~A. Saunders, Costas~D. Maranas, Nathan~E. Lewis,
  Thomas Sauter, Bernhard Palsson, Ines Thiele, and Ronan M.~T. Fleming.
\newblock {Creation and analysis of biochemical constraint-based models using
  the COBRA Toolbox v.3.0}.
\newblock \emph{Nature Protocols}, 14\penalty0 (3):\penalty0 639--702, 2019.
\newblock ISSN 1750-2799.
\newblock \doi{10.1038/s41596-018-0098-2}.

\bibitem[Karnopp et~al.(2012)Karnopp, Margolis, and Rosenberg]{KarMarRos12}
Dean~C Karnopp, Donald~L Margolis, and Ronald~C Rosenberg.
\newblock \emph{System Dynamics: Modeling, Simulation, and Control of
  Mechatronic Systems}.
\newblock John Wiley \& Sons, 5th edition, 2012.
\newblock ISBN 978-0470889084.

\bibitem[Karr et~al.(2012)Karr, Sanghvi, Macklin, Gutschow, Jacobs, Jr.,
  Assad-Garcia, Glass, and Covert]{KarSanMac12}
Jonathan~R. Karr, Jayodita~C. Sanghvi, Derek~N. Macklin, Miriam~V. Gutschow,
  Jared~M. Jacobs, Benjamin~Bolival Jr., Nacyra Assad-Garcia, John~I. Glass,
  and Markus~W. Covert.
\newblock A whole-cell computational model predicts phenotype from genotype.
\newblock \emph{Cell}, 150\penalty0 (2):\penalty0 389 -- 401, 2012.
\newblock ISSN 0092-8674.
\newblock \doi{10.1016/j.cell.2012.05.044}.

\bibitem[Klipp et~al.(2016)Klipp, Liebermeister, Wierling, and
  Kowald]{KliLieWie16}
Edda Klipp, Wolfram Liebermeister, Christoph Wierling, and Axel Kowald.
\newblock \emph{{Systems Biology: a Textbook}}.
\newblock Wiley-VCH, Weinheim, Germany, 2nd edition, 2016.

\bibitem[Lane(2014)]{Lan14}
Nick Lane.
\newblock Bioenergetic constraints on the evolution of complex life.
\newblock \emph{Cold Spring Harbor Perspectives in Biology}, 6\penalty0 (5),
  2014.
\newblock \doi{10.1101/cshperspect.a015982}.

\bibitem[Lane(2018)]{Lan18}
Nick Lane.
\newblock Hot mitochondria?
\newblock \emph{PLOS Biology}, 16\penalty0 (1):\penalty0 1--6, 01 2018.
\newblock \doi{10.1371/journal.pbio.2005113}.

\bibitem[Lark et~al.(2016)Lark, Torres, Lin, Ryan, Anderson, and
  Neufer]{LarTorLin16}
Daniel~S. Lark, Maria~J. Torres, Chien-Te Lin, Terence~E. Ryan, Ethan~J.
  Anderson, and P.~Darrell Neufer.
\newblock Direct real-time quantification of mitochondrial oxidative
  phosphorylation efficiency in permeabilized skeletal muscle myofibers.
\newblock \emph{American Journal of Physiology - Cell Physiology}, 311\penalty0
  (2):\penalty0 C239--C245, 2016.
\newblock ISSN 0363-6143.
\newblock \doi{10.1152/ajpcell.00124.2016}.

\bibitem[Lopaschuk and Dhalla(2014)]{LopDha14}
Gary~D. Lopaschuk and Naranjan~S. Dhalla, editors.
\newblock \emph{Cardiac Energy Metabolism in Health and Disease}.
\newblock Springer New York, New York, NY, 2014.
\newblock ISBN 978-1-4939-1227-8.
\newblock \doi{10.1007/978-1-4939-1227-8}.

\bibitem[Martin et~al.(2014)Martin, Sousa, and Lane]{MarSouLan14}
William~F. Martin, Filipa~L. Sousa, and Nick Lane.
\newblock Energy at life's origin.
\newblock \emph{Science}, 344\penalty0 (6188):\penalty0 1092--1093, 2014.
\newblock ISSN 0036-8075.
\newblock \doi{10.1126/science.1251653}.

\bibitem[{Medley} et~al.(2016){Medley}, {Goldberg}, and {Karr}]{MedGolKar16}
J.~K. {Medley}, A.~P. {Goldberg}, and J.~R. {Karr}.
\newblock Guidelines for reproducibly building and simulating systems biology
  models.
\newblock \emph{IEEE Transactions on Biomedical Engineering}, 63\penalty0
  (10):\penalty0 2015--2020, Oct 2016.
\newblock ISSN 0018-9294.
\newblock \doi{10.1109/TBME.2016.2591960}.

\bibitem[Neal et~al.(2016)Neal, Carlson, Thompson, James, Kim, Tran, Crampin,
  Cook, and Gennari]{NeaCarTho16}
Maxwell~L. Neal, Brian~E. Carlson, Christopher~T. Thompson, Ryan~C. James,
  Karam~G. Kim, Kenneth Tran, Edmund~J. Crampin, Daniel~L. Cook, and John~H.
  Gennari.
\newblock Semantics-based composition of integrated cardiomyocyte models
  motivated by real-world use cases.
\newblock \emph{PLoS ONE}, 10\penalty0 (12):\penalty0 1--18, 12 2016.
\newblock \doi{10.1371/journal.pone.0145621}.

\bibitem[Niebel et~al.(2019)Niebel, Leupold, and Heinemann]{NieLeuHei19}
Bastian Niebel, Simeon Leupold, and Matthias Heinemann.
\newblock An upper limit on {Gibbs} energy dissipation governs cellular
  metabolism.
\newblock \emph{Nature Metabolism}, 1\penalty0 (1):\penalty0 125--132, 2019.
\newblock ISSN 2522-5812.
\newblock \doi{10.1038/s42255-018-0006-7}.

\bibitem[Niven(2016)]{Niv16}
Jeremy~E Niven.
\newblock Neuronal energy consumption: biophysics, efficiency and evolution.
\newblock \emph{Current Opinion in Neurobiology}, 41:\penalty0 129 -- 135,
  2016.
\newblock ISSN 0959-4388.
\newblock \doi{10.1016/j.conb.2016.09.004}.

\bibitem[Niven and Laughlin(2008)]{NivLau08}
Jeremy~E. Niven and Simon~B. Laughlin.
\newblock Energy limitation as a selective pressure on the evolution of sensory
  systems.
\newblock \emph{Journal of Experimental Biology}, 211\penalty0 (11):\penalty0
  1792--1804, 2008.
\newblock ISSN 0022-0949.
\newblock \doi{10.1242/jeb.017574}.

\bibitem[{Noor}(2018)]{Noo18}
Elad {Noor}.
\newblock {Removing both Internal and Unrealistic Energy-Generating Cycles in
  Flux Balance Analysis}.
\newblock \emph{arXiv e-prints}, art. arXiv:1803.04999, Mar 2018.

\bibitem[Noor et~al.(2014)Noor, Bar-Even, Flamholz, Reznik, Liebermeister, and
  Milo]{NooBarFla14}
Elad Noor, Arren Bar-Even, Avi Flamholz, Ed~Reznik, Wolfram Liebermeister, and
  Ron Milo.
\newblock Pathway thermodynamics highlights kinetic obstacles in central
  metabolism.
\newblock \emph{PLOS Computational Biology}, 10\penalty0 (2):\penalty0 1--12,
  02 2014.
\newblock \doi{10.1371/journal.pcbi.1003483}.

\bibitem[Orth et~al.(2010{\natexlab{a}})Orth, Fleming, and
  Palsson]{OrtFlePal10}
J.~Orth, R.~Fleming, and B.~Palsson.
\newblock Reconstruction and use of microbial metabolic networks: the core
  escherichia coli metabolic model as an educational guide.
\newblock \emph{EcoSal Plus}, 2010{\natexlab{a}}.
\newblock \doi{10.1128/ecosalplus.10.2.1}.

\bibitem[Orth et~al.(2010{\natexlab{b}})Orth, Thiele, and Palsson]{OrtThiPal10}
Jeffrey~D. Orth, Ines Thiele, and Bernhard~O. Palsson.
\newblock What is flux balance analysis?
\newblock \emph{Nat Biotech}, 28:\penalty0 245--248, March 2010{\natexlab{b}}.
\newblock ISSN 1087-0156.
\newblock \doi{10.1038/nbt.1614}.

\bibitem[Orth et~al.(2011)Orth, Conrad, Na, Lerman, Nam, Feist, and
  Palsson]{OrtConNa11}
Jeffrey~D Orth, Tom~M Conrad, Jessica Na, Joshua~A Lerman, Hojung Nam, Adam~M
  Feist, and Bernhard~{\O} Palsson.
\newblock A comprehensive genome-scale reconstruction of escherichia coli
  metabolism{\textemdash}2011.
\newblock \emph{Molecular Systems Biology}, 7\penalty0 (1), 2011.
\newblock ISSN 1744-4292.
\newblock \doi{10.1038/msb.2011.65}.

\bibitem[Oster et~al.(1971)Oster, Perelson, and Katchalsky]{OstPerKat71}
George Oster, Alan Perelson, and Aharon Katchalsky.
\newblock Network thermodynamics.
\newblock \emph{Nature}, 234:\penalty0 393--399, December 1971.
\newblock \doi{10.1038/234393a0}.

\bibitem[Oster et~al.(1973)Oster, Perelson, and Katchalsky]{OstPerKat73}
George~F. Oster, Alan~S. Perelson, and Aharon Katchalsky.
\newblock Network thermodynamics: dynamic modelling of biophysical systems.
\newblock \emph{Quarterly Reviews of Biophysics}, 6\penalty0 (01):\penalty0
  1--134, 1973.
\newblock \doi{10.1017/S0033583500000081}.

\bibitem[Palsson(2006)]{Pal06}
Bernhard Palsson.
\newblock \emph{Systems biology: properties of reconstructed networks}.
\newblock Cambridge University Press, 2006.
\newblock ISBN 0521859034.

\bibitem[Palsson(2011)]{Pal11}
Bernhard Palsson.
\newblock \emph{Systems Biology: Simulation of Dynamic Network States}.
\newblock Cambridge University Press, 2011.

\bibitem[Palsson(2015)]{Pal15}
Bernhard Palsson.
\newblock \emph{{Systems Biology: Constraint-Based Reconstruction and
  Analysis}}.
\newblock {Cambridge University Press}, Cambridge, 2015.

\bibitem[{Pan} et~al.(2017){Pan}, {Gawthrop}, {Cursons}, {Tran}, and
  {Crampin}]{PanGawCurTraCra17X}
M.~{Pan}, P.~J. {Gawthrop}, J.~{Cursons}, K.~{Tran}, and E.~J. {Crampin}.
\newblock {The cardiac Na$^+$/K$^+$ ATPase: An updated, thermodynamically
  consistent model}.
\newblock Submitted, November 2017.

\bibitem[Pan et~al.(2018{\natexlab{a}})Pan, Gawthrop, Tran, Cursons, and
  Crampin]{PanGawTra18}
Michael Pan, Peter~J. Gawthrop, Kenneth Tran, Joseph Cursons, and Edmund~J.
  Crampin.
\newblock Bond graph modelling of the~cardiac action potential: implications
  for drift and non-unique steady states.
\newblock \emph{Proceedings of the Royal Society of London A: Mathematical,
  Physical and Engineering Sciences}, 474\penalty0 (2214), 2018{\natexlab{a}}.
\newblock ISSN 1364-5021.
\newblock \doi{10.1098/rspa.2018.0106}.
\newblock Available at {arXiv:1802.04548}.

\bibitem[Pan et~al.(2018{\natexlab{b}})Pan, Gawthrop, Tran, Cursons, and
  Crampin]{PanGawTra18a}
Michael Pan, Peter~J. Gawthrop, Kenneth Tran, Joseph Cursons, and Edmund~J.
  Crampin.
\newblock A thermodynamic framework for modelling membrane transporters.
\newblock \emph{Journal of Theoretical Biology}, 2018{\natexlab{b}}.
\newblock ISSN 0022-5193.
\newblock \doi{10.1016/j.jtbi.2018.09.034}.

\bibitem[Pan et~al.(2019)Pan, Gawthrop, Tran, Cursons, and
  Crampin]{PanGawTra19}
Michael Pan, Peter~J. Gawthrop, Kenneth Tran, Joseph Cursons, and Edmund~J.
  Crampin.
\newblock A thermodynamic framework for modelling membrane transporters.
\newblock \emph{Journal of Theoretical Biology}, 481:\penalty0 10 -- 23, 2019.
\newblock ISSN 0022-5193.
\newblock \doi{10.1016/j.jtbi.2018.09.034}.
\newblock Available at {arXiv:1806.04341}.

\bibitem[Park et~al.(2016)Park, Rubin, Xu, Amador-Noguez, Fan, Shlomi, and
  Rabinowitz]{ParRubXu16}
Junyoung~O. Park, Sara~A. Rubin, Yi-Fan Xu, Daniel Amador-Noguez, Jing Fan,
  Tomer Shlomi, and Joshua~D. Rabinowitz.
\newblock Metabolite concentrations, fluxes and free energies imply efficient
  enzyme usage.
\newblock \emph{Nat Chem Biol}, 12\penalty0 (7):\penalty0 482--489, Jul 2016.
\newblock ISSN 1552-4450.
\newblock \doi{10.1038/nchembio.2077}.

\bibitem[Paynter(1961)]{Pay61}
H.~M. Paynter.
\newblock \emph{{Analysis and Design of Engineering Systems}}.
\newblock MIT Press, Cambridge, Mass., 1961.

\bibitem[Polettini and Esposito(2014)]{PolEsp14}
Matteo Polettini and Massimiliano Esposito.
\newblock {Irreversible thermodynamics of open chemical networks. I. Emergent
  cycles and broken conservation laws}.
\newblock \emph{The Journal of Chemical Physics}, 141\penalty0 (2):\penalty0
  024117, 2014.
\newblock \doi{10.1063/1.4886396}.

\bibitem[Qian et~al.(2003)Qian, Beard, and Liang]{QiaBeaLia03}
Hong Qian, Daniel~A. Beard, and Shou-dan Liang.
\newblock Stoichiometric network theory for nonequilibrium biochemical systems.
\newblock \emph{European Journal of Biochemistry}, 270\penalty0 (3):\penalty0
  415--421, 2003.
\newblock ISSN 1432-1033.
\newblock \doi{10.1046/j.1432-1033.2003.03357.x}.

\bibitem[Schilling et~al.(2000)Schilling, Letscher, and Palsson]{SchLetPal00}
Christophe~H. Schilling, David Letscher, and Bernhard Palsson.
\newblock Theory for the systemic definition of metabolic pathways and their
  use in interpreting metabolic function from a pathway-oriented perspective.
\newblock \emph{Journal of Theoretical Biology}, 203\penalty0 (3):\penalty0 229
  -- 248, 2000.
\newblock ISSN 0022-5193.
\newblock \doi{10.1006/jtbi.2000.1073}.

\bibitem[Smith et~al.(2005)Smith, Barclay, and Loiselle]{SmiBarLoi05}
Nicholas~P. Smith, Christopher~J. Barclay, and Denis~S. Loiselle.
\newblock The efficiency of muscle contraction.
\newblock \emph{Progress in Biophysics and Molecular Biology}, 88\penalty0
  (1):\penalty0 1 -- 58, 2005.
\newblock ISSN 0079-6107.
\newblock \doi{10.1016/j.pbiomolbio.2003.11.014}.

\bibitem[Smith and Crampin(2004)]{SmiCra04}
N.P. Smith and E.J. Crampin.
\newblock Development of models of active ion transport for whole-cell
  modelling: cardiac sodium-potassium pump as a case study.
\newblock \emph{Progress in Biophysics and Molecular Biology}, 85\penalty0
  (2-3):\penalty0 387 -- 405, 2004.
\newblock \doi{10.1016/j.pbiomolbio.2004.01.010}.

\bibitem[Sousa et~al.(2013)Sousa, Thiergart, Landan, Nelson-Sathi, Pereira,
  Allen, Lane, and Martin]{SouThiLan13}
Filipa~L. Sousa, Thorsten Thiergart, Giddy Landan, Shijulal Nelson-Sathi,
  In{\^e}s A.~C. Pereira, John~F. Allen, Nick Lane, and William~F. Martin.
\newblock Early bioenergetic evolution.
\newblock \emph{Philosophical Transactions of the Royal Society of London B:
  Biological Sciences}, 368\penalty0 (1622), 2013.
\newblock ISSN 0962-8436.
\newblock \doi{10.1098/rstb.2013.0088}.

\bibitem[Swainston et~al.(2016)Swainston, Smallbone, Hefzi, Dobson, Brewer,
  Hanscho, Zielinski, Ang, Gardiner, Gutierrez, Kyriakopoulos, Lakshmanan, Li,
  Liu, Mart{\'i}nez, Orellana, Quek, Thomas, Zanghellini, Borth, Lee, Nielsen,
  Kell, Lewis, and Mendes]{SwaSmaHef16}
Neil Swainston, Kieran Smallbone, Hooman Hefzi, Paul~D. Dobson, Judy Brewer,
  Michael Hanscho, Daniel~C. Zielinski, Kok~Siong Ang, Natalie~J. Gardiner,
  Jahir~M. Gutierrez, Sarantos Kyriakopoulos, Meiyappan Lakshmanan, Shangzhong
  Li, Joanne~K. Liu, Veronica~S. Mart{\'i}nez, Camila~A. Orellana, Lake-Ee
  Quek, Alex Thomas, Juergen Zanghellini, Nicole Borth, Dong-Yup Lee, Lars~K.
  Nielsen, Douglas~B. Kell, Nathan~E. Lewis, and Pedro Mendes.
\newblock Recon 2.2: from reconstruction to model of human metabolism.
\newblock \emph{Metabolomics}, 12\penalty0 (7):\penalty0 109, Jun 2016.
\newblock ISSN 1573-3890.
\newblock \doi{10.1007/s11306-016-1051-4}.

\bibitem[Szigeti et~al.(2018)Szigeti, Roth, Sekar, Goldberg, Pochiraju, and
  Karr]{SziYosSek18}
Balázs Szigeti, Yosef~D. Roth, John~A.P. Sekar, Arthur~P. Goldberg, Saahith~C.
  Pochiraju, and Jonathan~R. Karr.
\newblock A blueprint for human whole-cell modeling.
\newblock \emph{Current Opinion in Systems Biology}, 7:\penalty0 8 -- 15, 2018.
\newblock ISSN 2452-3100.
\newblock \doi{10.1016/j.coisb.2017.10.005}.

\bibitem[Thiele et~al.(2013)Thiele, Swainston, Fleming, Hoppe, Sahoo, Aurich,
  Haraldsdottir, Mo, Rolfsson, Stobbe, Thorleifsson, Agren, Bolling, Bordel,
  Chavali, Dobson, Dunn, Endler, Hala, Hucka, Hull, Jameson, Jamshidi, Jonsson,
  Juty, Keating, Nookaew, Le~Novere, Malys, Mazein, Papin, Price, Selkov~Sr,
  Sigurdsson, Simeonidis, Sonnenschein, Smallbone, Sorokin, van Beek, Weichart,
  Goryanin, Nielsen, Westerhoff, Kell, Mendes, and Palsson]{ThiSwaFle13}
Ines Thiele, Neil Swainston, Ronan M.~T. Fleming, Andreas Hoppe, Swagatika
  Sahoo, Maike~K. Aurich, Hulda Haraldsdottir, Monica~L. Mo, Ottar Rolfsson,
  Miranda~D. Stobbe, Stefan~G. Thorleifsson, Rasmus Agren, Christian Bolling,
  Sergio Bordel, Arvind~K. Chavali, Paul Dobson, Warwick~B. Dunn, Lukas Endler,
  David Hala, Michael Hucka, Duncan Hull, Daniel Jameson, Neema Jamshidi,
  Jon~J. Jonsson, Nick Juty, Sarah Keating, Intawat Nookaew, Nicolas Le~Novere,
  Naglis Malys, Alexander Mazein, Jason~A. Papin, Nathan~D. Price, Evgeni
  Selkov~Sr, Martin~I. Sigurdsson, Evangelos Simeonidis, Nikolaus Sonnenschein,
  Kieran Smallbone, Anatoly Sorokin, Johannes H. G.~M. van Beek, Dieter
  Weichart, Igor Goryanin, Jens Nielsen, Hans~V. Westerhoff, Douglas~B. Kell,
  Pedro Mendes, and Bernhard~O. Palsson.
\newblock A community-driven global reconstruction of human metabolism.
\newblock \emph{Nat Biotech}, 31:\penalty0 419--425, May 2013.
\newblock ISSN 1087-0156.
\newblock \doi{10.1038/nbt.2488}.

\bibitem[Van~Rysselberghe(1958)]{Rys58}
Pierre Van~Rysselberghe.
\newblock Reaction rates and affinities.
\newblock \emph{The Journal of Chemical Physics}, 29\penalty0 (3):\penalty0
  640--642, 1958.
\newblock \doi{10.1063/1.1744552}.

\bibitem[Vander~Heiden et~al.(2009)Vander~Heiden, Cantley, and
  Thompson]{HeiCanTho09}
Matthew~G. Vander~Heiden, Lewis~C. Cantley, and Craig~B. Thompson.
\newblock Understanding the warburg effect: The metabolic requirements of cell
  proliferation.
\newblock \emph{Science}, 324\penalty0 (5930):\penalty0 1029--1033, 2009.
\newblock ISSN 0036-8075.
\newblock \doi{10.1126/science.1160809}.

\end{thebibliography}

\appendix

% \section{Bond graphs}
% \label{sec:bond-graphs-2}
% \begin{figure}[htbp]
%   \centering
%   \Fig{pAB_abg}{0.3}
%   \caption{Example: Parallel reaction \S~\ref{sec:exampl-parall-react}}
%   \label{fig:pAB}
% \end{figure}

% \begin{figure}[htbp]
%   \centering
%   \Fig{ABC_abg}{0.3}
%   \caption{Example: three-reaction cycle from \S~\ref{sec:exampl-three-react}}
%   \label{fig:ABC}
% \end{figure}

% \begin{figure}[htbp]
%   \centering
%   \Fig{Toy_abg}{0.3}
%   \caption{Illustrative example \cite{Noo18} from \S~\ref{sec:ilustr-example-citen}}
%   \label{fig:Toy_abg}
% \end{figure}

\section{Glycolysis \& Pentose Phosphate Pathways: Reactions}
\label{sec:glycolysis--pentose}

\begin{xalignat*}{2}
\ch{GLCD_E + PEP &<>[ GLCPTS ] G6P + PYR }&&(-26.81~\si{\kilo\joule\per\mol})\\
\ch{G6P &<>[ PGI ] F6P }&&(-1.60~\si{\kilo\joule\per\mol})\\
\ch{ATP + F6P &<>[ PFK ] ADP + FDP + H }&&(-24.71~\si{\kilo\joule\per\mol})\\
\ch{FDP &<>[ FBA ] DHAP + G3P }&&(-1.98~\si{\kilo\joule\per\mol})\\
\ch{DHAP &<>[ TPI ] G3P }&&(-0.79~\si{\kilo\joule\per\mol})\\
\ch{G3P + NAD + PI &<>[ GAPD ] 13DPG + H + NADH }&&(-1.32~\si{\kilo\joule\per\mol})\\
\ch{3PG + ATP &<>[ PGK ] 13DPG + ADP }&&(-1.42~\si{\kilo\joule\per\mol})\\
\ch{2PG &<>[ PGM ] 3PG }&&(-3.17~\si{\kilo\joule\per\mol})\\
\ch{2PG &<>[ ENO ] H2O + PEP }&&(-2.75~\si{\kilo\joule\per\mol})\\
\ch{ADP + H + PEP &<>[ PYK ] ATP + PYR }&&(-7.09~\si{\kilo\joule\per\mol})
\end{xalignat*}
\begin{xalignat*}{2}
\ch{G6P + NADP &<>[ G6PDH2R ] 6PGL + H + NADPH }&&(-0.96~\si{\kilo\joule\per\mol})\\
\ch{6PGL + H2O &<>[ PGL ] 6PGC + H }&&(-0.96~\si{\kilo\joule\per\mol})\\
\ch{6PGC + NADP &<>[ GND ] CO2 + NADPH + RU5PD }&&(-15.08~\si{\kilo\joule\per\mol})\\
\ch{RU5PD &<>[ RPI ] R5P }&&(-0.00~\si{\kilo\joule\per\mol})\\
\ch{E4P + XU5PD &<>[ TKT2 ] F6P + G3P }&&(-1.61~\si{\kilo\joule\per\mol})\\
\ch{G3P + S7P &<>[ TALA ] E4P + F6P }&&(-5.43~\si{\kilo\joule\per\mol})\\
\ch{R5P + XU5PD &<>[ TKT1 ] G3P + S7P }&&(-0.40~\si{\kilo\joule\per\mol})\\
\ch{RU5PD &<>[ RPE ] XU5PD }&&(-0.08~\si{\kilo\joule\per\mol})
\end{xalignat*}

\section{Modular representation of Metabolism: Reactions}
\label{sec:modul-repr-metab}

\subsection{Glycolysis}
\begin{align}
\ch{GLCD_E + PEP &<>[ GLCPTS ] G6P + PYR }\\
\ch{G6P &<>[ PGI ] F6P }\\
\ch{ATP + F6P &<>[ PFK ] ADP + FDP + H }\\
\ch{FDP + H2O &<>[ FBP ] F6P + PI }\\
\ch{FDP &<>[ FBA ] DHAP + G3P }\\
\ch{DHAP &<>[ TPI ] G3P }\\
\ch{G3P + NAD + PI &<>[ GAPD ] 13DPG + H + NADH }\\
\ch{3PG + ATP &<>[ PGK ] 13DPG + ADP }\\
\ch{2PG &<>[ PGM ] 3PG }\\
\ch{2PG &<>[ ENO ] H2O + PEP }
\end{align}

\begin{align}
\ch{ADP + H + PEP &<>[ PYK ] ATP + PYR }
\end{align}

\subsection{TCA cycle}
\begin{align}
\ch{COA + NAD + PYR &<>[ PDH ] ACCOA + CO2 + NADH }\\
\ch{ACCOA + H2O + OAA &<>[ CS ] CIT + COA + H }\\
\ch{CIT &<>[ ACONTA ] ACONC + H2O }\\
\ch{ACONC + H2O &<>[ ACONTB ] ICIT }\\
\ch{ICIT + NADP &<>[ ICDHYR ] AKG + CO2 + NADPH }\\
\ch{AKG + COA + NAD &<>[ AKGDH ] CO2 + NADH + SUCCOA }\\
\ch{ATP + COA + SUCC &<>[ SUCOAS ] ADP + PI + SUCCOA }\\
\ch{FUM + Q8H2 &<>[ FRD7 ] Q8 + SUCC }\\
\ch{Q8 + SUCC &<>[ SUCDI ] FUM + Q8H2 }\\
\ch{FUM + H2O &<>[ FUM ] MALL }
\end{align}

\begin{align}
\ch{MALL + NAD &<>[ MDH ] H + NADH + OAA }\\
\ch{NAD + NADPH &<>[ NADTRHD ] NADH + NADP }
\end{align}

\subsection{Electron Transport Chain}
\begin{align}
\ch{4 H + NADH + Q8 &<>[ NADH16 ] 3 H_E + NAD + Q8H2 }\\
\ch{4 H + O2 + 2 Q8H2 &<>[ CYTBD ] 2 H2O + 4 H_E + 2 Q8 }
\end{align}

\subsection{ATPase}
\begin{align}
\ch{ADP + 4 H_E + PI &<>[ ATPS4R ] ATP + 3 H + H2O }
\end{align}

\end{document}